\documentclass[prd,aps,nofootinbib,twocolumn,floatfix,10 pt]{revtex4}
\usepackage{amssymb}
\usepackage{amsmath,graphicx,color,epsfig}

\begin{document}

\title{Fourth-generation SM imprints in $B\to K^{\ast}\ell^{+}\ell^{-}$ decays with polarized $K^{\ast}$}
\author{Aqeel Ahmed,\footnote{aqeel@ncp.edu.pk} Ishtiaq Ahmed,\footnote{ishtiaq@ncp.edu.pk} M. Jamil Aslam,\footnote{jamil@ncp.edu.pk} M. Junaid,\footnote{mjunaid@ncp.edu.pk} M. Ali Paracha,\footnote{ali@ncp.edu.pk} and Abdur Rehman\footnote{rehman@ncp.edu.pk}}
\affiliation{National Centre for Physics and Physics Department,\\
Quaid-i-Azam University, Islamabad 45320, Pakistan.}
\date{\today}

\begin{abstract}
The implication of the fourth-generation quarks in the $B\to
K^{\ast}\ell^{+}\ell^{-}$ ($\ell=\mu,\tau$) decays, when $K^{\ast}$
meson is longitudinally or transversely polarized, is presented. In
this context, the dependence of the branching ratio with polarized
$K^{\ast}$ and the helicity fractions ($f_{L,T}$) of $K^{\ast}$
meson are studied. It is observed that the polarized branching
ratios as well as helicity fractions are sensitive to the NP
parameters, especially when the final state leptons are tauons.
Hence the measurements of these observables at LHC can serve as a
good tool to investigate the indirect searches of new physics beyond
the Standard Model.

\end{abstract}

\maketitle




\section{Introduction}

\label{intro}

 It is well known that the Standard Model (SM) with single Higgs
 boson is the simplest one and has been tested with great
 precision. Despite its many  successes it has some theoretical shortcomings which
 preclude to recognize the SM as a fundamental theory. For example the SM doesn't
 addresses the following issues (i) hierarchy puzzle (ii) origin of
 mass spectrum (iii) Why only three generations of quarks and
 leptons? (iv) Neutrinos are massless but experiments have shown that the neutrinos have non-zero mass.

 The above mentioned issues indicate that there must be a new
 physics (NP) beyond the SM. Various extensions of the SM are
 motivated to understand some of the above mentioned problems. These
 extensions are two Higgs doublet models (2HDM), minimal
 supersymmetric SM (MSSM), universal extra dimension model (UED) and
 Standard model with fourth generation (SM4)etc. SM4 implying a
 fourth family of quarks and leptons seems to most economical in
 number of additional particles and simpler in the sense that it
 does not introduce any new operators. Interest in SM4 was fairly high in the 1980s until the electroweak precision
data seemed to rule it out. The other reason which increases the
interest in the fourth generation was the measurement of the number
of light neutrinos at the $Z$ pole that showed only three light
neutrinos could exist. However, the discovery of neutrino
oscillations suggested the possibility of a mass scale beyond the
SM, and the models with the sufficiently massive neutrino became
acceptable \cite{Pree}. Though the early study of the EW precision
measurements ruled out a fourth generation \cite{PDG}, however it
was subsequently pointed out \cite{Maltoni} that if the fourth
generation masses are not degenerate, then the EW precision data do
not prohibit the fourth generation \cite{Kible}. Therefore, the SM
can be simply extended with a sequential  as four quark and four
lepton left handed doublets and corresponding right handed singlets.

The possible sequential fourth generation may play an important role
in understanding the well known problem of CP violation and flavor
structure of standard theory \cite{3, 4, 5, 6, 7, 8, 9}, electroweak
symmetry breaking \cite{10, 11, 12, 13}, hierarchies of fermion mass
and mixing angle in quark/lepton sectors \cite{14, 15, 16, 17, 18}.
A thorough discussion on the theoretical and experimental aspects of
the fourth generation can be found in ref. \cite{19}.

It is necessary to mention here that the SM4 particles are heavy in
nature,  consequently they are hard to produce in the accelerators.
Therefore, we have to go for some alternate scenarios where we can
find their influence at low energies. In this regard, the Flavor
Changing Neutral Current (FCNC) transitions provide an ideal
plateform to establish new physics (NP). This is because of the fact
that FCNC transitions are not allowed at tree level in the SM and
are allowed at loop level through GIM mechanism which can get
contributions of NP from newly proposed particles via loop diagrams.
Among different FCNC transitions the one $b\rightarrow s$ transition
plays a pivotal role to perform efficient tests of NP scenarios
\cite{25, 26,
27, 28, 29, 30, 31, 32, 33}. It is also the fact that $CP$ violation in $%
b\rightarrow s $ transitions is predicted to be very small in the
SM, thus, any experimental evidence for sizable $CP$ violating
effects in the $B$ system would clearly point towards a NP scenario.
The FCNC transitions in SM4 contains much fewer parameters and has a
possibility of having simultaneously sizeable effects in $K$ and $B$
mesons system compared to other NP models.

The exploration of Physics beyond the Standard model through various
inclusive $B$ meson decays like $B\to X_{s,d}\ell^{+}\ell^{-}$ and
their corresponding exclusive processes, $B\to M \ell^{+}\ell^{-}$
with $M = K,K^{\ast},K_{1},\rho $ etc have been done in literature
\cite{bst, 63, zp, colangelo, pbr, aqeel}. These studies showed that
the above mentioned inclusive and exclusive decays of $B$ meson are
very sensitive to the flavor structure of the Standard Model and
provide a windowpane for any NP model. There are two different ways
to incorporate the NP effects in the rare decays, one through the
modification in Wilson coefficients and the other through new
operators which are absent in the Standard Model. It is necessary to
mention here the FCNC decay modes like $B\to X_{s}\ell^{+}\ell^{-}$,
$B\to K^{\ast}\ell^{+}\ell^{-}$ and $B\to K \ell^{+}\ell^{-}$ which
are also useful in the determination of precise values of
$C_{7}^{eff}$, $C_{9}^{eff}$ and $C_{10}^{eff}$ Wilson coefficients
as well as the sign of $C_{7}^{eff}.$ In particular these decay
modes involved observables which can distinguish between the various
extensions of Standard Model.

The observables like branching ratio, forward-backward
asymmetry,lepton polarization asymmetries and helicity fractions of
final state mesons for the semileptonic $B$ decays are greatly
influenced under different scenarios beyond the Standard Model.
Therefore, the precise measurement of these observables will play an
important role in the indirect searches of NP including SM4. In this
work we study the physical observables such as branching ratio and
helicity fractions when $K^{\ast}$ meson is longitudinally and
transversely polarized for the decays $B\to
K^{\ast}\ell^{+}\ell^{-}$ in SM4. The longitudinal helicity fraction
$f_{L}$ has been measured for the $K^{\ast}$ meson by
LHCb\cite{LHCb}, CDF\cite{CDF},Belle \cite{Belle} and Babar
\cite{babar} collaborations for the decay channel $B\to
K^{\ast}\mu^{+}\mu^{-}$. The recent results are quite intriguing
especially that from LHCb and CDF collaborations. In this respect it
is appropriate to look for the above mentioned observables which can
be tested experimentally to pin down the status of SM4. In exclusive
decays the main job is to calculate the form factors, for our
analysis we borrow the light cone QCD sum rules form factors
(LCSR)\cite{Ball}.

We organize the manuscript as follows: In sec. \ref{tf}, we fill our
toolbox with the theoretical framework needed to study the said
process in the fourth-generation SM. In Sec. \ref{pheno}, we discuss
the phenomenology of the polarized branching ratios and helicity
fractions of $K^{\ast}$ meson in $B\rightarrow
K^{\ast}\ell^{+}\ell^{-}$ in detail. We give the numerical analysis
of our observables and discuss the sensitivity of these observables
with the NP scenarios. We summarize the main points of our findings
in Sec. \ref{conc}.

\section{Theoretical Toolbox}\label{tf}
At quark level the decay $B\rightarrow K^{\ast} \ell^{+}\ell^{-}$
$(\ell=\mu,\tau)$ is governed by the transition $b\rightarrow s
\ell^{+} \ell^{-}$ for which the effective Hamiltonian can be
written as
\begin{equation}
H_{eff}=-\frac{4G_{F}}{\sqrt{2}}V_{tb}V^{\ast }_{ts}{\sum\limits_{i=1}^{10}}%
C_{i}({\mu })O_{i}({\mu }),  \label{effective hamiltonian 1}
\end{equation}%
where $O_{i}({\mu })$ $(i=1,\ldots ,10)$ are the four-quark operators and $%
C_{i}({\mu })$ are the corresponding Wilson\ coefficients at the
energy scale ${\mu }$ and the explicit expressions of these in the
SM at NLO and NNLO are given in \cite{Buchalla, Buras, Ball, Ali,
Kim, Kruger,Grinstein, Cella,
Bobeth, Asatrian, Misiak, Huber}. The operators responsible for $%
B\rightarrow K^{\ast }\ell^{+}\ell^{-}$ are $O_{7}$, $O_{9}$ and
$O_{10}$ and their form is given by
\begin{eqnarray}
O_{7} &=&\frac{e^{2}}{16\pi ^{2}}m_{b}\left( \bar{s}\sigma _{\mu \nu
}P_{R}b\right) F^{\mu \nu },\,  \notag \\
O_{9} &=&\frac{e^{2}}{16\pi ^{2}}(\bar{s}\gamma _{\mu
}P_{L}b)(\bar{l}\gamma
^{\mu }l),\,  \label{op-form} \\
O_{10} &=&\frac{e^{2}}{16\pi ^{2}}(\bar{s}\gamma _{\mu }P_{L}b)(\bar{l}%
\gamma ^{\mu }\gamma _{5}l),  \notag
\end{eqnarray}%
with $P_{L,R}=\left( 1\mp \gamma _{5}\right) /2$.

In terms of the above Hamiltonian, the free quark decay amplitude for $%
b\rightarrow s$ $\ell^{+}\ell^{-}$ in SM4 can be derived as:
\begin{align}
&\mathcal{M}(b \rightarrow s \ell^{+}\ell^{-})
= -\frac{G_{F}\alpha}{\sqrt{2}%
\pi } V_{tb}V_{ts}^{\ast } \bigg\{ C_{9}^{eff}(\bar{s}\gamma _{\mu
}L b)(\bar{\ell} \gamma ^{\mu
}\ell)\notag\\
&+C_{10}(\bar{s}\gamma _{\mu }L b)(\bar{\ell}\gamma ^{\mu }\gamma
_{5}\ell)
-2m_{b}C_{7}^{eff}(\bar{s}i\sigma _{\mu \nu }\frac{q^{\nu }}{q^{2}}R b)(\bar{ \ell%
}\gamma ^{\mu }\ell) \bigg\}\label{quark-amplitude}
\end{align}
where $q^{2}$ is the square of momentum transfer. The operator
$O_{10}$ can not be induced by the insertion of four-quark operators
because of the absence of the $Z$-boson in the effective theory.
Therefore, the Wilson coefficient $C_{10}$ does not renormalize
under QCD corrections and hence it is independent on the energy
scale. In addition to this, the above quark level decay amplitude
can receive contributions from the matrix element of four-quark
operators, $\sum_{i=1}^{6}\langle \ell^{+}\ell^{-}s|O_{i}|b\rangle
$,
which are usually absorbed into the effective Wilson coefficient $%
C_{9}^{SM}(\mu )$ and can usually be called $C_{9}^{eff}$, that can
be found in \cite{Ball, aqeel}

The sequential fourth generation model with an additional up-type
quark $t^{\prime }$ and down-type quark $b^{\prime }$ , a heavy
charged lepton $\tau^{\prime}$ and an associated neutrino
$\nu^{\prime}$ is a simple and non-supersymmetric extension of the
SM, and as such does not add any new dynamics to the SM. Being a
simple extension of the SM it retains all the properties of the SM
where the new top quark $t^{\prime }$ like the other up-type quarks,
contributes to $b\to s$ transition at the loop level. Therefore, the
effect of fourth generation displays itself by
changing the values of Wilson coefficients $C_{7}\left( \mu \right) $, $%
C_{9}\left( \mu \right) $ and $C_{10}$ via the virtual exchange of
fourth generation up-type quark $t^{\prime }$ which then take the
form;
\begin{equation}
\lambda _{t}C_{i}\rightarrow \lambda _{t}C_{i}^{SM}+\lambda
_{t^{\prime }}C_{i}^{new},  \label{wilson-modified}
\end{equation}%
where $\lambda _{f}=V_{fb}^{\ast }V_{fs}$ and the explicit forms of the $%
C_{i}$'s can be obtained from the corresponding expressions of the
Wilson coefficients in the SM by substituting $m_{t}\rightarrow
m_{t^{\prime }}$. By adding an extra family of quarks, the CKM
matrix of the SM is extended by another row and column which now
becomes $4\times 4$. The unitarity of which
leads to%
\begin{equation*}
\lambda _{u}+\lambda _{c}+\lambda _{t}+\lambda _{t^{\prime }}=0.
\end{equation*}%
Since $\lambda _{u}=V_{ub}^{\ast }V_{us}$ has a very small value
compared to the others, therefore, we will ignore it. Then $\lambda
_{t}\approx -\lambda
_{c}-\lambda _{t^{\prime }}$ and from Eq. (\ref{wilson-modified}) we have%
\begin{equation}
\lambda _{t}C_{i}^{SM}+\lambda _{t^{\prime }}C_{i}^{new}=-\lambda
_{c}C_{i}^{SM}+\lambda _{t^{\prime }}\left(
C_{i}^{new}-C_{i}^{SM}\right) . \label{wilson-modified1}
\end{equation}%
One can clearly see that under $\lambda _{t^{\prime }}\rightarrow 0$ or $%
m_{t^{\prime }}\rightarrow m_{t}$ the term $\lambda _{t^{\prime
}}\left( C_{i}^{new}-C_{i}^{SM}\right) $ vanishes which is the
requirement of GIM mechanism. Taking the contribution of the
$t^{\prime }$ quark in the loop
the Wilson coefficients $C_{i}$'s can be written in the following form%
\begin{eqnarray}
C_{7}^{tot}\left( \mu \right) &=&C_{7}^{eff SM}\left( \mu \right)
+\frac{\lambda
_{t^{\prime }}}{\lambda _{t}}C_{7}^{new}\left( \mu \right) ,  \notag \\
C_{9}^{tot}\left( \mu \right) &=&C_{9}^{eff SM}\left( \mu \right)
+\frac{\lambda _{t^{\prime }}}{\lambda _{t}}C_{9}^{new}\left( \mu
\right) ,
\label{wilson-tot} \\
C_{10}^{tot}\left( \mu \right) &=&C_{10}^{SM}\left( \mu \right)+\frac{\lambda _{t^{\prime }}}{\lambda _{t}}%
C_{10}^{new}\left( \mu \right),  \notag
\end{eqnarray}%
where we factored out $\lambda _{t}=V_{tb}^{\ast }V_{ts}$ term in
the effective Hamiltonian given in Eq. (\ref{effective hamiltonian
1}) and the
last term in these expressions corresponds to the contribution of the $%
t^{\prime }$ quark to the Wilson Coefficients. $\lambda _{t^{\prime
}}$ can be parameterized as:
\begin{equation}
\lambda _{t^{\prime }}=\left\vert V_{t^{\prime }b}^{\ast
}V_{t^{\prime }s}\right\vert e^{i\phi _{sb}}
\end{equation}%
where $\phi_{sb}$ is the new $CP$ odd phase.

\subsection{Parametrization of the Matrix Elements and Form Factors}\label{ff}

The exclusive $B\rightarrow K^{\ast }\ell^{+}\ell^{-}$ decay
involves the hadronic matrix elements which can be obtained by
sandwiching the quark
level operators given in Eq. (\ref{quark-amplitude}) between initial state $%
B$ meson and final state $K^{\ast }$ meson. These can be
parameterized in terms of the form factors which are the scalar
functions of the square of the four momentum
transfer($q^{2}=(p-k)^{2}).$ The non vanishing matrix elements for
the process $B\rightarrow K^{\ast }$ can be
parameterized in terms of the seven form factors as follows%
\begin{align}
\left\langle K^{\ast }(k,\varepsilon )\left\vert \bar{s}\gamma _{\mu
}b\right\vert B(p)\right\rangle &=\epsilon _{\mu \nu \alpha \beta
}\varepsilon
^{\ast \nu }p^{\alpha }k^{\beta } \frac{2A_{V}(q^{2})}{M_{B}+M_{K^{\ast}}} \label{10} \\
\left\langle K^{\ast }(k,\varepsilon )\left\vert \bar{s}\gamma _{\mu
}\gamma _{5}b\right\vert B(p)\right\rangle &=i\varepsilon ^{\ast
}_{\mu }\left(
M_{B}+M_{K^{\ast }}\right) A_{1}(q^{2}) \notag \\
&-i(\varepsilon ^{\ast }\cdot q)\left( p+k\right) _{\mu } \frac{A_{2}\left( q^{2}\right) }{M_{B}+M_{K^{\ast }}} \notag \\
&-i2(\varepsilon ^{\ast }\cdot q)q_{\mu } M_{K^{\ast
}}\frac{A_{3}\left( q^{2}\right)-A_{0}\left( q^{2}\right) }{q^{2}}
\label{11}
\end{align}%
where $p$ is the momentum of $B$ meson and, $\varepsilon $ $(k)$ are
the polarization vector (momentum) of the final state $K^{\ast }$
meson. In Eq. (\ref{11}) we use the following exact relation
\begin{equation}
A_{3}(q^{2})=\frac{M_{B}+M_{K^{\ast }}}{2M_{K^{\ast }}}A_{1}(q^{2})-\frac{M_{B}-M_{K^{\ast }}}{%
2M_{K^{\ast }}}A_{2}(q^{2})  \label{tf8}
\end{equation}%
with
\begin{equation*}
A_{3}(0)=A_{0}(0)
\end{equation*}%
and
\begin{equation}
\left\langle K^{\ast }(k,\varepsilon )\left\vert \partial_{\mu}
\gamma^{\mu}\right\vert B(p)\right\rangle =2M_{K^{\ast}}\varepsilon
^{\ast \mu }p_{\mu} A_{0}(q^{2}) \label{eom}
\end{equation}

In addition to the above form factors there are some penguin form
factors, which we can write as
\begin{align}
&\left\langle K^{\ast }(k,\varepsilon )\left\vert \bar{s}\sigma
_{\mu \nu }q^{\nu }b\right\vert B(p)\right\rangle =2i\epsilon _{\mu
\nu \alpha \beta }\varepsilon ^{\ast \nu }p^{\alpha }k^{\beta }
T_{1}(q^{2}) \label{13a}
\\
&\left\langle K^{\ast }(k,\varepsilon )\left\vert \bar{s}\sigma
_{\mu \nu
}q^{\nu }\gamma ^{5}b\right\vert B(p)\right\rangle \notag \\
&=\bigg\{ \left( M_{B}^{2}-M_{K^{\ast }}^{2}\right) \varepsilon
_{\mu }^{\ast
}-(\varepsilon ^{\ast }\cdot q)(p+k)_{\mu }\bigg\} T_{2}(q^{2})  \notag \\
&+(\varepsilon ^{\ast }\cdot q)\bigg\{ q_{\mu }-\frac{q^{2}}{%
M_{B}^{2}-M_{K^{\ast }}^{2}}(p+k)_{\mu }\bigg\} T_{3}(q^{2}).
\label{13b}
\end{align}%
The form factors $A_{V}\left( q^{2}\right)$, $A_{1}\left(
q^{2}\right),$ $A_{2}\left( q^{2}\right),$ $A_{3}\left(
q^{2}\right),$ $A_{0}\left( q^{2}\right)$,
$T_{1}\left(q^{2}\right)$, $T_{2}\left( q^{2}\right)$, $T_{3}\left(
q^{2}\right)$ are the non-perturbative quantities and to calculate
them one has to rely on some non-perturbative approaches. To study the physical observables in the high $q^{2}$ bin, we take the form factors calculated
 in the framework of LCSR \cite{Ball}. The dependence of the form factors on square of the momentum transfer $(q^{2})$ can be written as

\begin{equation}
F\left( q^{2}\right) =F\left( 0\right) Exp\bigg[c_{1}\frac{q^{2}}{M_{B}^{2}%
}+c_{2} \frac{q^{4}}{M_{B}^{4}}\bigg].  \label{ff-param}
\end{equation}%
where the values of the parameters $F\left( 0\right) $, $c_{1} $ and $%
c_{2} $ are given in Table \ref{formfactor}.

\begin{table}[tbh]
\centering \caption{$B\rightarrow K^{\ast }$ form factors
corresponding to penguin contributions in the light cone QCD Sum
Rules(LCSR). $F(0)$ denotes the value of form factors at $q^{2}=0$
while $c_{1}$ and $c_{2}$ are the parameters in the parametrization
shown in Eq. (\ref{ff-param}) \cite{Ball}.}
\label{formfactor}%
\begin{tabular}{cccc}
\hline\hline
 $F(q^{2})$ & $\hspace{1cm}F(0)\hspace{1cm}$ & $\hspace{1cm}c_{1}\hspace{1cm}$ & $\hspace{1cm}c_{2}\hspace{0.5cm}$ \\ \hline
 $A_{V}\left( q^{2}\right) $ & $0.457^{+0.091}_{-0.058}$ & $1.482$ & $1.015$ \\ \hline
 $A_{1}(q^{2})$ & $0.337^{+0.048}_{-0.043}$ & $
0.602$ & $0.258$ \\ \hline
 $A_{2}(q^{2})$ & $0.282^{+0.038}_{-0.036}$ & $
1.172$ & $0.567$ \\ \hline
 $A_{0}(q^{2})$ & $0.471^{+0.227}_{-0.059}$ & $
1.505$ & $0.710$ \\ \hline
 $T_{1}(q^{2})$ & $0.379^{+0.058}_{-0.045}$ & $
1.519$ & $1.030$ \\ \hline
 $T_{2}(q^{2})$ & $0.379^{+0.058}_{-0.045}$ & $
0.517$ & $0.426$ \\ \hline
 $T_{3}(q^{2})$ & $0.260^{+0.035}_{-0.026}$ & $
1.129$ & $1.128$ \\ \hline\hline
\end{tabular}
\end{table}

Now in terms of these form factors and from Eq.
(\ref{quark-amplitude}) it is straightforward to write the penguin
amplitude as
\begin{equation*}
\mathcal{M}=-\frac{G_{F}\alpha }{2\sqrt{2}\pi }%
V_{tb}V_{ts}^{\ast }\left[ {\cal T}_{\mu }^{1}(\bar{l}\gamma ^{\mu
}l)+{\cal T}_{\mu }^{2}\left( \bar{l}\gamma ^{\mu }\gamma
^{5}l\right) \right]
\end{equation*}%
where%
\begin{align}
{\cal T}_{\mu }^{1} &=f_{1}(q^{2})\epsilon _{\mu \nu \alpha \beta
}\varepsilon ^{\ast \nu }p^{\alpha }k^{\beta
}-if_{2}(q^{2})\varepsilon _{\mu }^{\ast
}+if_{3}(q^{2})(\varepsilon ^{\ast }\cdot q)P_{\mu }  \label{60} \\
{\cal T}_{\mu }^{2} &=f_{4}(q^{2})\epsilon _{\mu \nu \alpha \beta
}\varepsilon ^{\ast \nu }p^{\alpha }k^{\beta
}-if_{5}(q^{2})\varepsilon _{\mu }^{\ast
}\notag \\
&+if_{6}(q^{2})(\varepsilon ^{\ast }\cdot q)P_{\mu }
+if_{0}(q^{2})(\varepsilon ^{\ast }\cdot q)q_{\mu } \label{61}
\end{align}

The functions $f_{0}$ to $f_{6}$ in Eq. (\ref{60}) and Eq.
(\ref{61}) are known as auxiliary functions, which contain both long
distance (form factors) and short distance (Wilson coefficients)
effects and these can be written as
\begin{subequations}
\begin{align}
f_{1}(q^{2}) &=4(m_{b}+m_{s})\frac{C_{7}^{tot}}{q^{2}}%
T_{1}(q^{2})+2C_{9}^{tot}\frac{A_{V}(q^{2})}{M_{B}+M_{K^{\ast }}}\label{62a} \\
f_{2}(q^{2})
&=\frac{2C_{7}^{tot}}{q^{2}}(m_{b}-m_{s})T_{2}(q^{2})\left(
M_{B}^{2}-M_{K^{\ast }}^{2}\right) \notag\\
&+C_{9}^{tot}A_{1}(q^{2})\left(
M_{B}+M_{K^{\ast }}\right)   \\
f_{3}(q^{2}) &= 4\frac{C_{7}^{tot}}{q^{2}}(m_{b}-m_{s})\left(
T_{2}(q^{2})+q^{2}\frac{T_{3}(q^{2})}{\left( M_{B}^{2}-M_{K^{\ast
}}^{2}\right) }\right) \notag\\
&+2C_{9}^{tot}\frac{A_{2}(q^{2})}{M_{B}+M_{K^{%
\ast }}}  \\
f_{4}(q^{2}) &=C_{10}^{tot}\frac{2A_{V}(q^{2})}{M_{B}+M_{K^{\ast }}}
 \\
f_{5}(q^{2}) &=C_{10}^{tot}A_{1}(q^{2})\left( M_{B}+M_{K^{\ast
}}\right)    \\
f_{6}(q^{2}) &=C_{10}^{tot}\frac{A_{2}(q^{2})}{M_{B}+M_{K^{\ast }}}
\\
f_{0}(q^{2})
&=C_{10}^{tot}\frac{A_{3}(q^{2})-A_{0}(q^{2})}{M_{B}+M_{K^{\ast }}}
\label{62g}
\end{align}%
\end{subequations}

\section{Phenomenological Observables}\label{pheno}

\subsection{Polarized Branching Ratio}

The explicit expression of the differential decay rate for
$B\rightarrow K^{\ast}\ell^{+}\ell^{-}$, when the $K^{\ast}$ meson
is polarized, can be written in terms of longitudinal $\Gamma_{L}$
and transverse components $\Gamma_{T}$ as \cite{pbr}
\begin{eqnarray}
\frac{d\Gamma _{L}(q^{2})}{dq^{2}} &=&\frac{G_{F}^{2}\left\vert V_{tb}V_{ts}^{\ast }\right\vert ^{2}\alpha ^{2}}{2^{11}\pi ^{5}}\frac{u(q^{2})}{M_{B}^{3}}\times \frac{1}{3}{\cal A}_{L}  \label{65d} \\
\frac{d\Gamma _{\pm }(q^{2})}{dq^{2}} &=&\frac{%
G_{F}^{2}\left\vert V_{tb}V_{ts}^{\ast }\right\vert ^{2}\alpha ^{2}}{%
2^{11}\pi ^{5}}\frac{u(q^{2})}{M_{B}^{3}}\times \frac{4}{3}{\cal A}_{\pm }  \label{65g}\\
\frac{d\Gamma _{T}(q^{2})}{dq^{2}} &=&\frac{d\Gamma _{+
}(q^{2})}{dq^{2}} +\frac{d\Gamma _{- }(q^{2})}{dq^{2}}
\end{eqnarray}%
where are the polarized branching ratios can be defined as
\begin{equation}
BR_{L,T}=\frac{\int_{q_{\min}^{2}}^{q_{\max}^{2}}\frac{d\Gamma_{L,T}(q^{2})}{dq^{2}}dq^{2}}{\Gamma_{total}}.\\
\end{equation}
The kinematical variables used in above equations are defined as

\begin{equation}
u(q^{2})\equiv
\sqrt{\lambda\left(1-\frac{4m_{\ell}^{2}}{q^{2}}\right)},\label{uq}\\
\end{equation}
with
\begin{eqnarray}
\lambda &\equiv
&\lambda\left(m_{B}^{2},m_{K^{\ast}}^{2},q^{2}\right)\notag\\
&=&m_{B}^{4}+m_{K^{\ast}}^{4}+q^{4}-2m_{K^{\ast}}^{2}m_{B}^{2}-2q^{2}m_{B}^{2}-2q^{2}m_{K^{\ast}}^{2}.\notag\\
&&\label{lambda}
\end{eqnarray}
The different functions appearing in Eqs. (\ref{65d}-\ref{65g}) can
be expressed in terms of auxiliary functions (cf. Eqs.
(\ref{62a}-\ref{62g})) as
\begin{align}
{\cal A}_{L} &=\frac{1}{q^{2}M_{K^{\ast }}^{2}}\bigg[24\left\vert
f_{0}(q^{2})\right\vert ^{2}m^{2}M_{K^{\ast }}^{2}\lambda
+\notag \\
&(2m^{2}+q^{2})\left\vert (M_{B}^{2}-M_{K^{\ast
}}^{2}-q^{2})f_{2}(q^{2})+\lambda f_{3}(q^{2})\right\vert ^{2}  \notag \\
&+(q^{2}-4m^{2})\left\vert (M_{B}^{2}-M_{K^{\ast
}}^{2}-q^{2})f_{5}(q^{2})+\lambda f_{6}(q^{2})\right\vert ^{2}\bigg]  \label{65j}\\
{\cal A}_{\pm} &=(q^{2}-4m^{2})\left\vert f_{5}(q^{2})\mp
\sqrt{\lambda }f_{4}(q^{2})\right\vert ^{2}\notag \\
&+\left( q^{2}+2m^{2}\right) \left\vert f_{2}(q^{2})\mp\sqrt{\lambda
}f_{1}(q^{2})\right\vert ^{2} \label{65q}
\end{align}

\subsection{Helicity Fractions of $K^{\ast}$ Meson}

We now discuss helicity fractions of $K^{\ast }$ in $B\rightarrow
K^{\ast }\ell^{+}\ell^{-}$ which are interesting observables and are
as such independent of the uncertainties arising due to form factors
and other input parameters. The final state meson helicity fractions
were already discussed in literature for $B\rightarrow K^{\ast
}\left( K_{1}\right) \ell^{+}\ell^{-}$ decays
\cite{colangelo,aqeel,23,aa1}. The longitudinal helicity fraction
$f_{L}$ has been measured for the $K^{\ast }$ vector meson, by the
LHCb \cite{LHCb}, CDF \cite{CDF}, Belle \cite{Belle} and Babar
\cite{babar} collaborations for the decay $B\rightarrow K^{\ast
}\ell^{+}\ell^{-}(l=e,\mu )$ in the region of low momentum transfer
$( 0.1\leq q^{2}\leq 6 \text{ GeV}^{2})$ the results are given
below:
\begin{subequations}
\begin{eqnarray}
f_{L} &=&0.57_{-0.10}^{+0.11}\pm 0.03, \ \   \text{(LHCb)} \label{64a}\\
f_{L} &=&0.69_{-0.21}^{+0.19}\pm 0.08, \ \  \text{(CDF)} \label{64b}\\
f_{L} &=&0.67_{-0.23}^{+0.23}\pm 0.05, \ \   \text{(Belle)} \label{64c}\\
f_{L} &=&0.35_{-0.16}^{+0.16}\pm 0.04, \ \   \text{(Babar)}
\label{64d}
\end{eqnarray}%
\end{subequations}
while the SM average value of $f_{L}$ in  $0.1\leq q^{2}\leq 6
\text{ GeV}^{2})$ range is $f_{L}=0.65$ where the average value of
the helicity fractions is defined as:
\begin{equation}
\left\langle f_{L,T}(q^{2})\right\rangle =\frac{\int_{q_{\min
}^{2}}^{q_{\max
}^{2}}f_{L,T}(q^{2})\frac{dBR_{L,T}}{dq^{2}}dq^{2}}{\int_{q_{\min
}^{2}}^{q_{\max }^{2}}\frac{dBR_{L,T}}{dq^{2}}dq^{2}}.
\end{equation}

Finally the longitudinal and transverse helicity amplitude becomes
\begin{eqnarray}
f_{L}(q^{2}) &=&\frac{d\Gamma _{L}(q^{2})/dq^{2}}{d\Gamma
(q^{2})/dq^{2}}
\notag \\
f_{\pm }(q^{2}) &=&\frac{d\Gamma _{\pm }(q^{2})/dq^{2}}{d\Gamma
(q^{2})/dq^{2}}  \notag \\
f_{T}(q^{2}) &=&f_{+}(q^{2})+f_{-}(q^{2})
\end{eqnarray}%
so that \ the sum of the longitudinal and transverse helicity
amplitudes is equal to one i.e. $f_{L}(q^{2})+f_{T}(q^{2})=1$ for
each value of $q^{2}$ \cite{colangelo}.

\subsection{Numerical Work and Discussion}

In this section we analyze the impact of SM4 on the observables like
longitudinal  branching ratio ($BR_{L}$),transverse  branching ratio
($BR_{T}$) and helicity fractions of $K^{\ast}$ for $B\to
K^{\ast}\ell^{+}\ell^{-}$ $(\ell=\mu, \tau)$ decays. In the
numerical calculation of the said physical observables, the LCSR
form factor are used which are given in Table I and the values of
Wilson Coefficients are given in Table \ref{wc table}, while the
other input parameters are collected in Table \ref{input}.
\begin{table}[ht]
\centering \caption{Default values of input parameters used in the
calculations \cite{pdg}}
\begin{tabular}{c}
\hline\hline
$m_{B}=5.28$ GeV, $m_{b}=4.28$ GeV, $m_{\mu}=0.105$ GeV,\\
$m_{\tau}=1.77$ GeV, $f_{B}=0.25$ GeV,
$|V_{tb}V_{ts}^{\ast}|=45\times
10^{-3}$,\\ $\alpha^{-1}=137$, $G_{F}=1.17\times 10^{-5}$ GeV$^{-2}$,\\
$\tau_{B}=1.54\times 10^{-12}$ sec, $m_{K^{\ast}}=0.892$ GeV.\\
\hline\hline
\end{tabular}\label{input}
\end{table}
\begin{table*}[ht]
\centering \caption{The Wilson coefficients $C_{i}^{\mu}$ at the
scale $\mu\sim m_{b}$ in the SM \cite{Ball}.}
\begin{tabular}{cccccccccc}
\hline\hline
$C_{1}$&$C_{2}$&$C_{3}$&$C_{4}$&$C_{5}$&$C_{6}$&$C_{7}$&$C_{9}$&$C_{10}$
\\ \hline
 \ \ \ 1.107 \ \ \ & \ \ \ -0.248 \ \ \ & \ \ \ -0.011 \ \ \ & \ \ \ -0.026 \ \ \ & \ \ \ -0.007 \ \ \ & \ \ \ -0.031 \ \ \ & \ \ \ -0.313 \ \ \ & \ \ \ 4.344 \ \ \ & \ \ \ -4.669 \ \ \ \\
\hline\hline
\end{tabular}
\label{wc table}
\end{table*}

\subsubsection{NP in Polarized branching ratios $BR_{L}$ and $BR_{T}$}

\begin{itemize}

\item In Figs. 1 and 2 we have plotted the $BR_{L}$ and $BR_{T}$ against the $q^{2}(GeV^{2})$ for $\mu$ and $\tau$ as final state leptons for the said decay. In these graphs we set the value
$\phi_{sb}=90^{\circ}$ and vary the values of $m_{t'}$ and
$|V_{t'b}V_{t's}|$ such that they lies well within the constraints
obtained from different $B$-meson decays \cite{ASoni}. These graphs
indicate that both $BR_{L}$ and $BR_{T}$ are increasing function of
the SM4 parameters. One can also see that at the minimum values of
the SM4 parameters, the NP effects are masked by the uncertainties,
especially for the tauons as final state leptons. However, when we
set the maximum values of these parameters, the increment in both
the $BR_{L}$ and $BR_{T}$, lie well above the uncertainties in the
SM values as shown in the Figs. 1 and 2.

\item To see the explicit dependence on the SM4
parameters we have integrated out $BR_{L}$ and $BR_{T}$ over $q^{2}$
and have drawn both of them against the $m_{t'}, |V_{t'b}V_{t's}|$
and $\phi_{sb}$ in Figs. 3-5. In Fig. 3 we have plotted $BR_{L}$ and
$BR_{T}$ vs $m_{t'}$ where $\phi_{sb}$ is set to be $90^{\circ}$ and
choose the three different values of $|V_{t'b}V_{t's}|$ i.e 0.005,
0.01 and 0.015. These graphs clearly depict that as the value of
$m_{t'}$ is increased the $BR_{L}$ and $BR_{T}$ are enhanced
accordingly. For the case of muons as a final state leptons (see
Figs. 3(a,b)) the increment in the $BR_{L}$ and $BR_{T}$ values, at
the maximum value of $m_{t'}=600 GeV$, is up to 5 times that of the
SM values while in the case of tauns which is presented in Figs.
3(c, d), the increment in the $BR_{L}$ and $BR_{T}$ values is
approximately 3 to 4 times larger than that of the SM values.

\item In Fig. 4, $BR_{L}$ and $BR_{T}$ are plotted as a function of
$|V_{t'b}V_{t's}|$ where three different curves correspond to the
three different values of $m_{t'}=300, 450, 600 GeV$ and
$\phi_{sb}=60^{\circ}, 90^{\circ}, 120^{\circ}$ as shown in the
graphs. From these graphs one can easily see that similar to the
case of $m_{t^{\prime}}$ the $BR_{L}$ and $BR_{T}$ is also an
increasing function of $|V_{t'b}V_{t's}|$.

\item To see how $BR_{L}$ and $BR_{T}$ are evolved due to the variation in the CKM4 phase
$\phi_{sb}$, we have plotted $BR_{L}$ and $BR_{T}$ vs $\phi_{sb}$ in
Fig. 5. It is noticed that in contrast to the previous two cases for
$m_{t'}$ and $|V_{t'b}V_{t's}|$ the $BR_{L}$ and $BR_{T}$ are
increasing when $\phi_{sb}$ is decreasing. It is easy to extract
from the graph that at $\phi_{sb}=60^{\circ}, m_{t^{\prime}}=600
GeV$ and $|V_{t'b}V_{t's}|=0.015$ the values of $BR_{L}$ and
$BR_{T}$ are about 6 to 7 times larger than that of their SM values
both for muons and tauons. Further more, the BR for each value of
$|V_{t'b}V_{t's}|$ decrease to almost half when $\phi _{sb}$ reaches
$120^\circ$ (Fig. 5).

\end{itemize}
\begin{figure*}[ht]
\begin{tabular}{cc}
\hspace{0.6cm}($\mathbf{a}$)&\hspace{1.2cm}($\mathbf{b}$)\\
\includegraphics[scale=0.5]{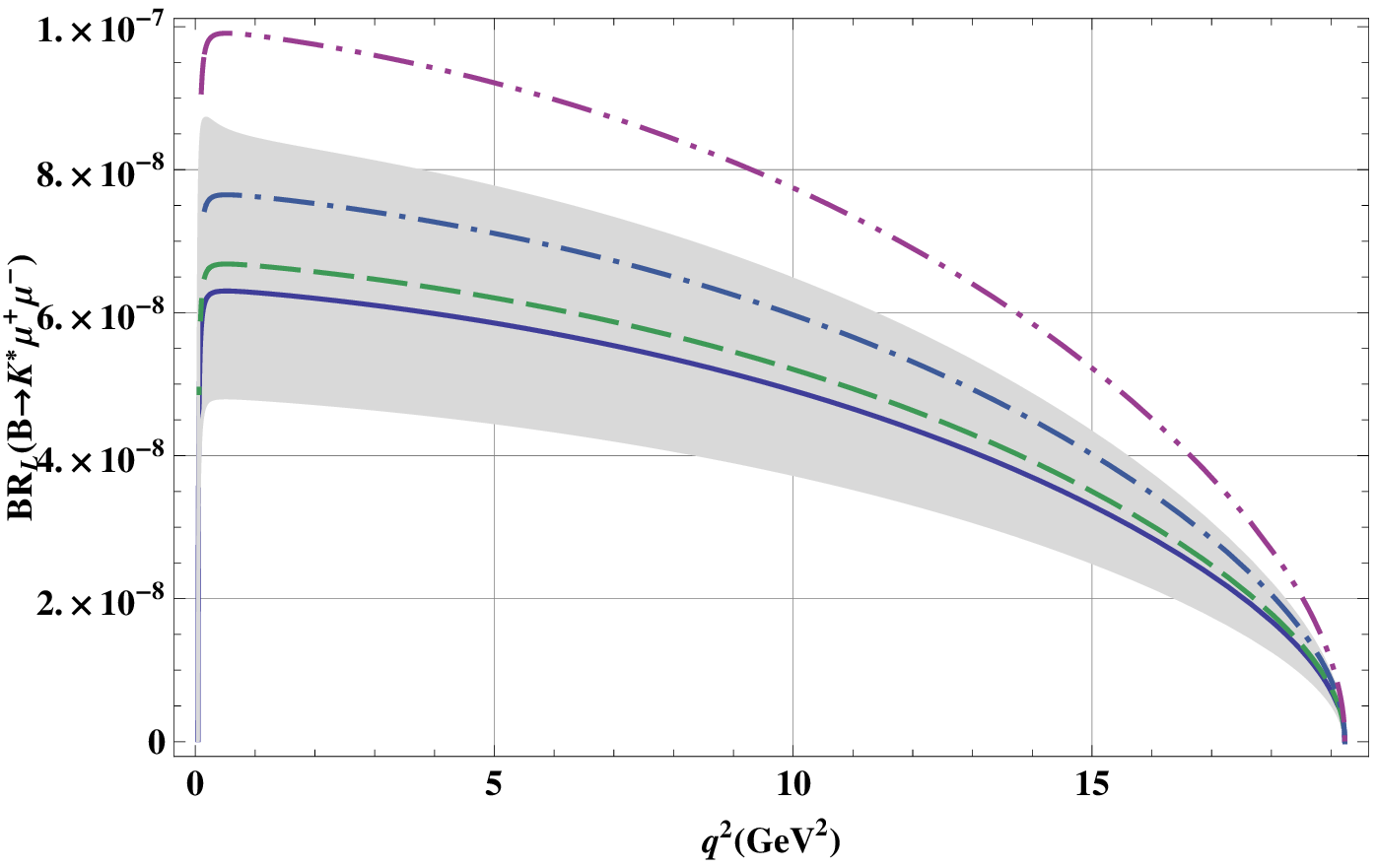} \ \ \
& \ \ \ \includegraphics[scale=0.5]{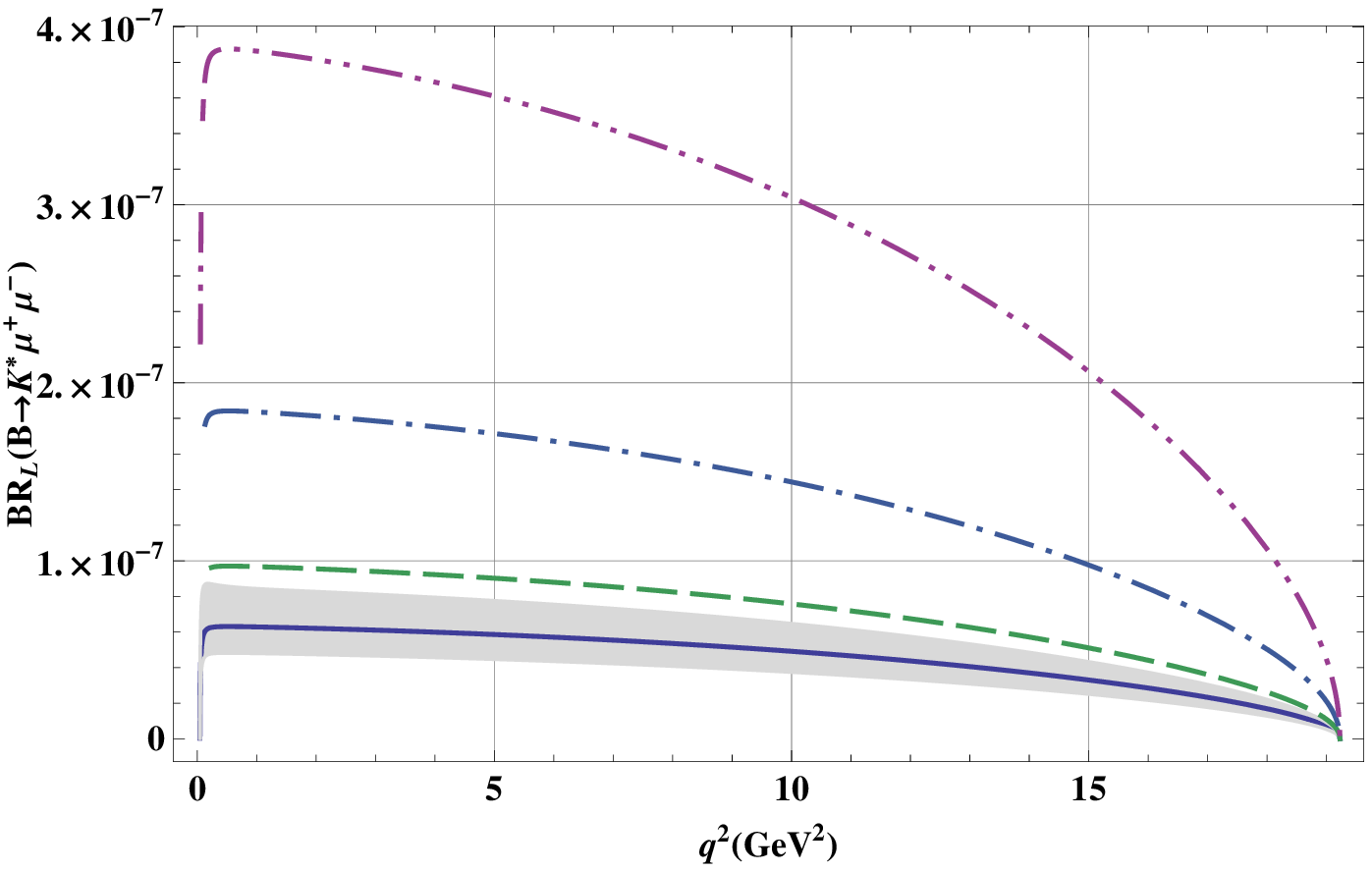}\\
\hspace{0.6cm}($\mathbf{c}$)&\hspace{1.2cm}($\mathbf{d}$)\\
\includegraphics[scale=0.5]{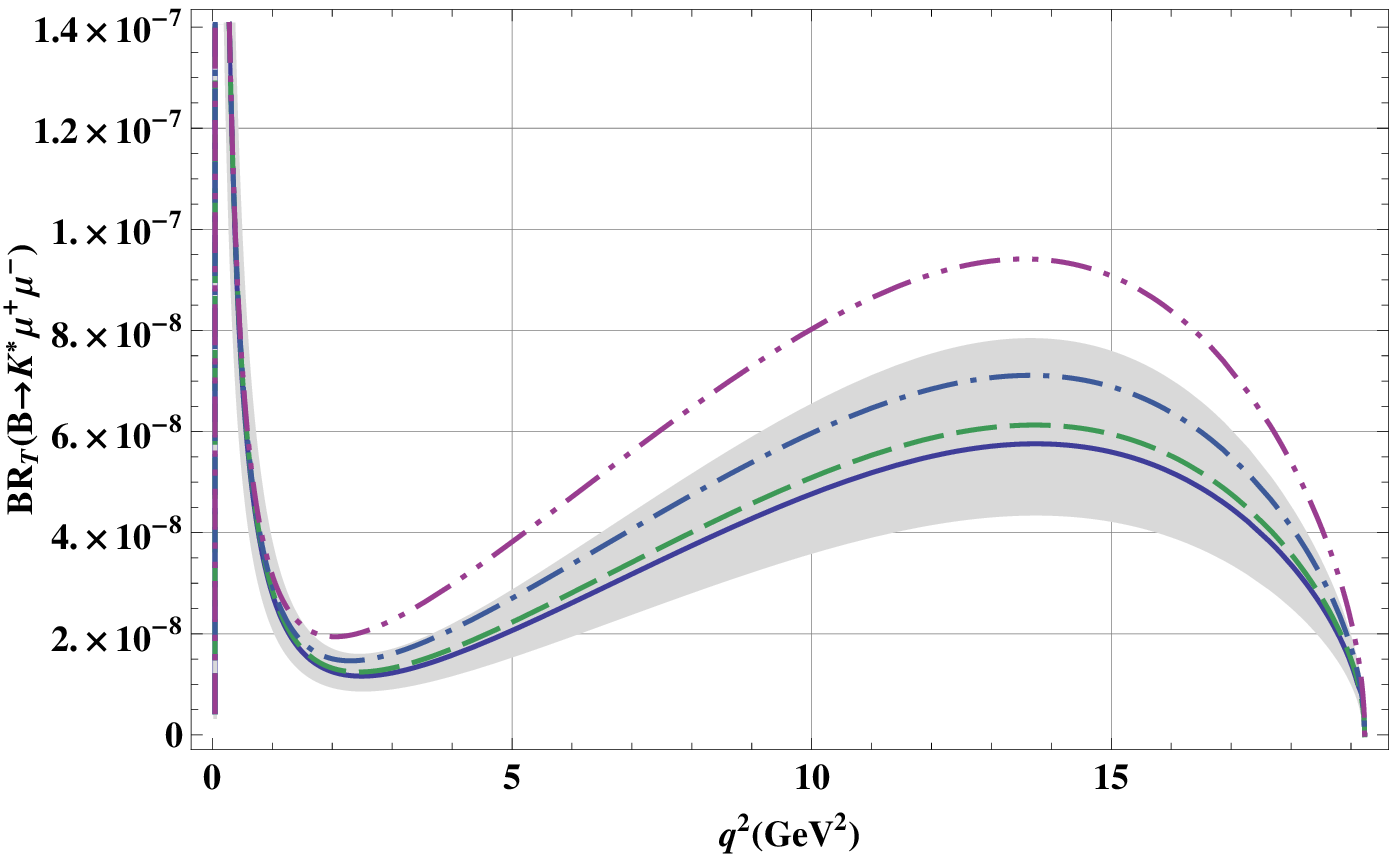} \ \ \
& \ \ \ \includegraphics[scale=0.5]{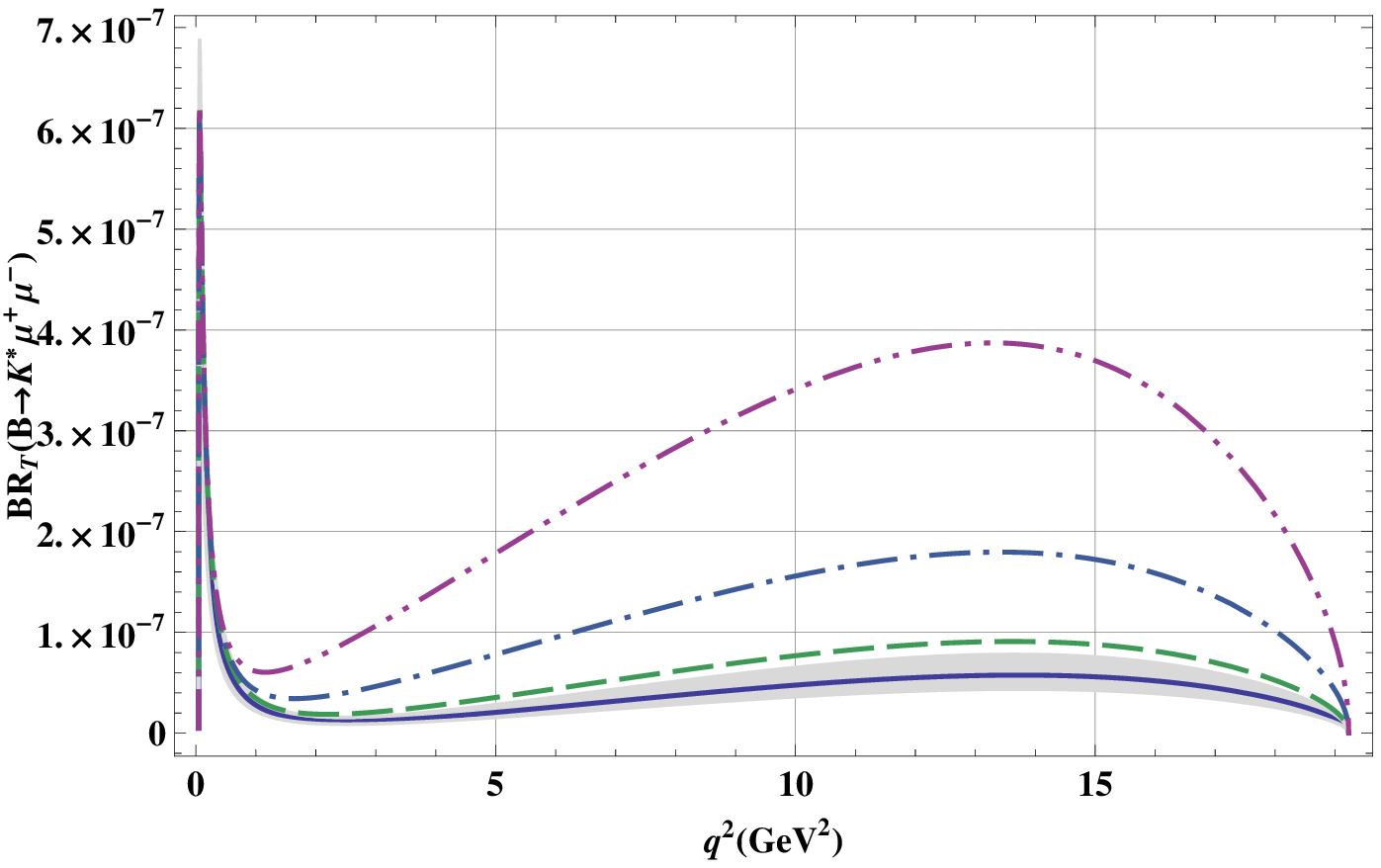}
\end{tabular}
\caption{The dependence of the longitudinal and transverse BR for
the decay $B\to K^*(892)\mu^{+}\mu^{-}$ on $q^{2}$ for different
values of $m_{t^{\prime}}$ and $\left\vert
V^{\ast}_{t^{\prime}b}V_{t^{\prime}s}\right\vert$. In all the
graphs, the solid line corresponds to the SM, dashed , dashed dot
and dashed double  corresponds to $m_{t^{\prime}}=$ 300 GeV, 450 GeV
and 600 GeV respectively. $\left\vert
V^{\ast}_{t^{\prime}b}V_{t^{\prime}s}\right\vert$ has the value
0.005 and 0.015 in $(a)$ and $(b)$ respectively.} \label{lpm}
\end{figure*}

\begin{figure*}[ht]
\begin{tabular}{cc}
\hspace{0.6cm}($\mathbf{a}$)&\hspace{1.2cm}($\mathbf{b}$)\\
\includegraphics[scale=0.5]{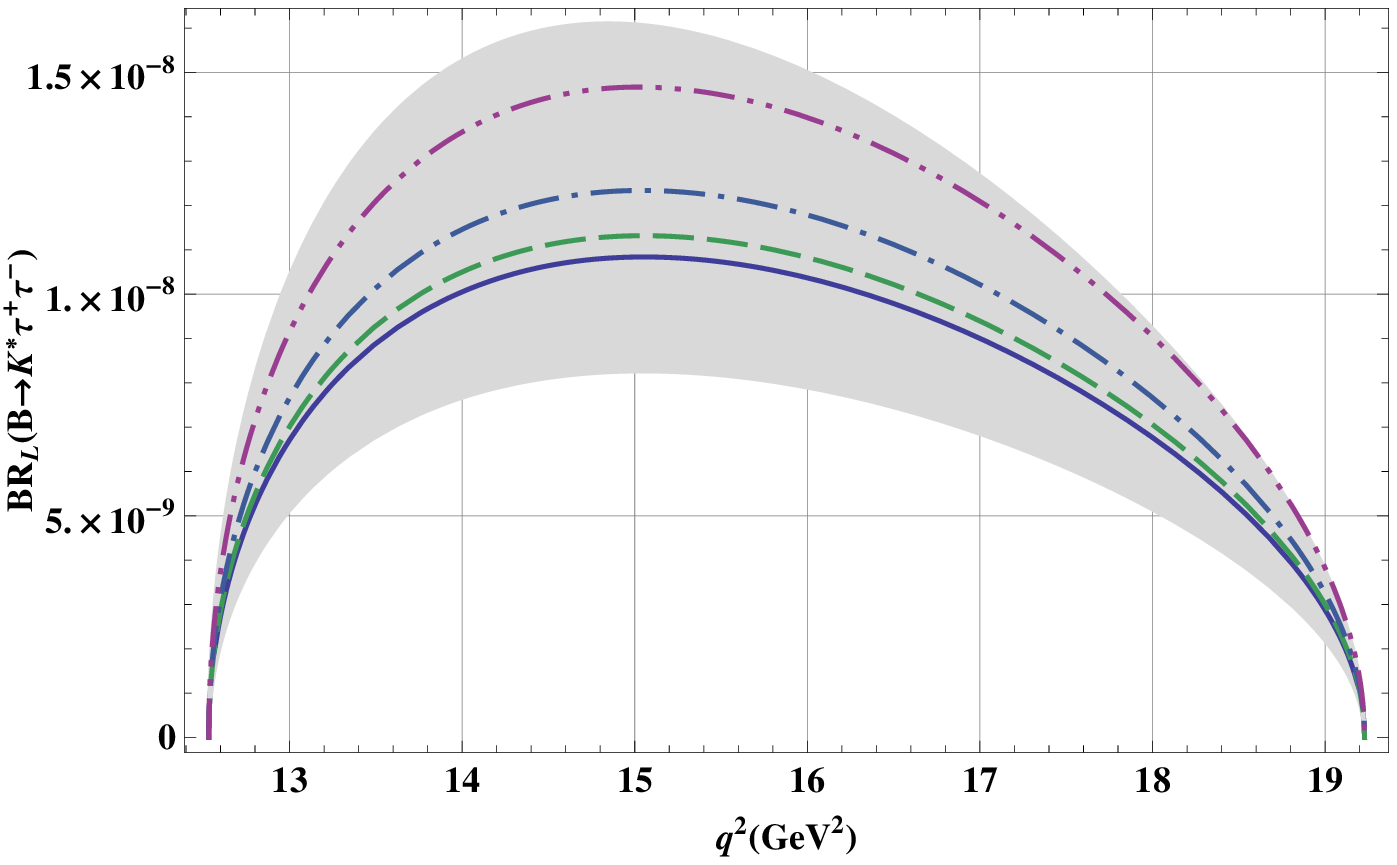} \ \ \
& \ \ \ \includegraphics[scale=0.5]{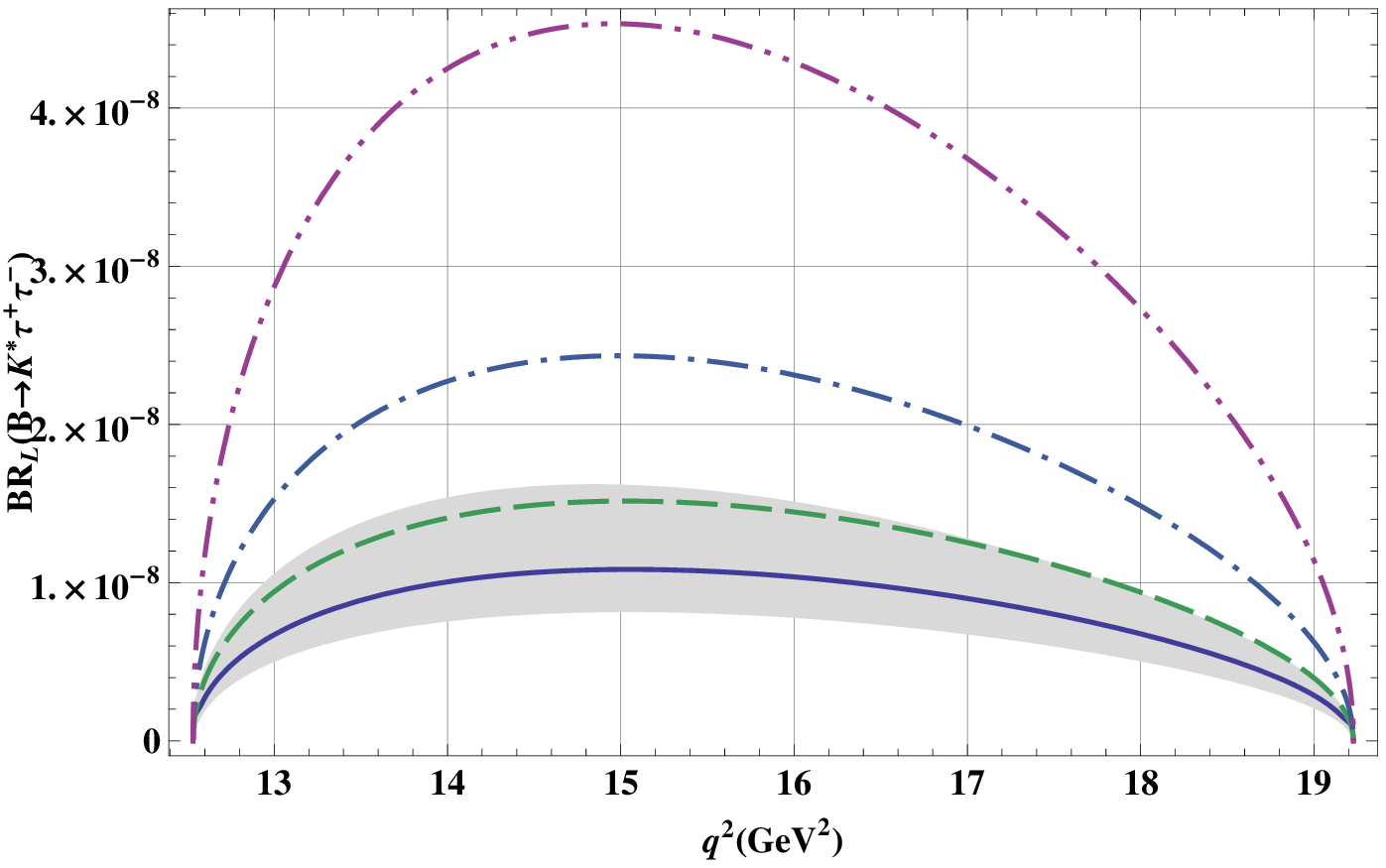}\\
\hspace{0.6cm}($\mathbf{c}$)&\hspace{1.2cm}($\mathbf{d}$)\\
\includegraphics[scale=0.5]{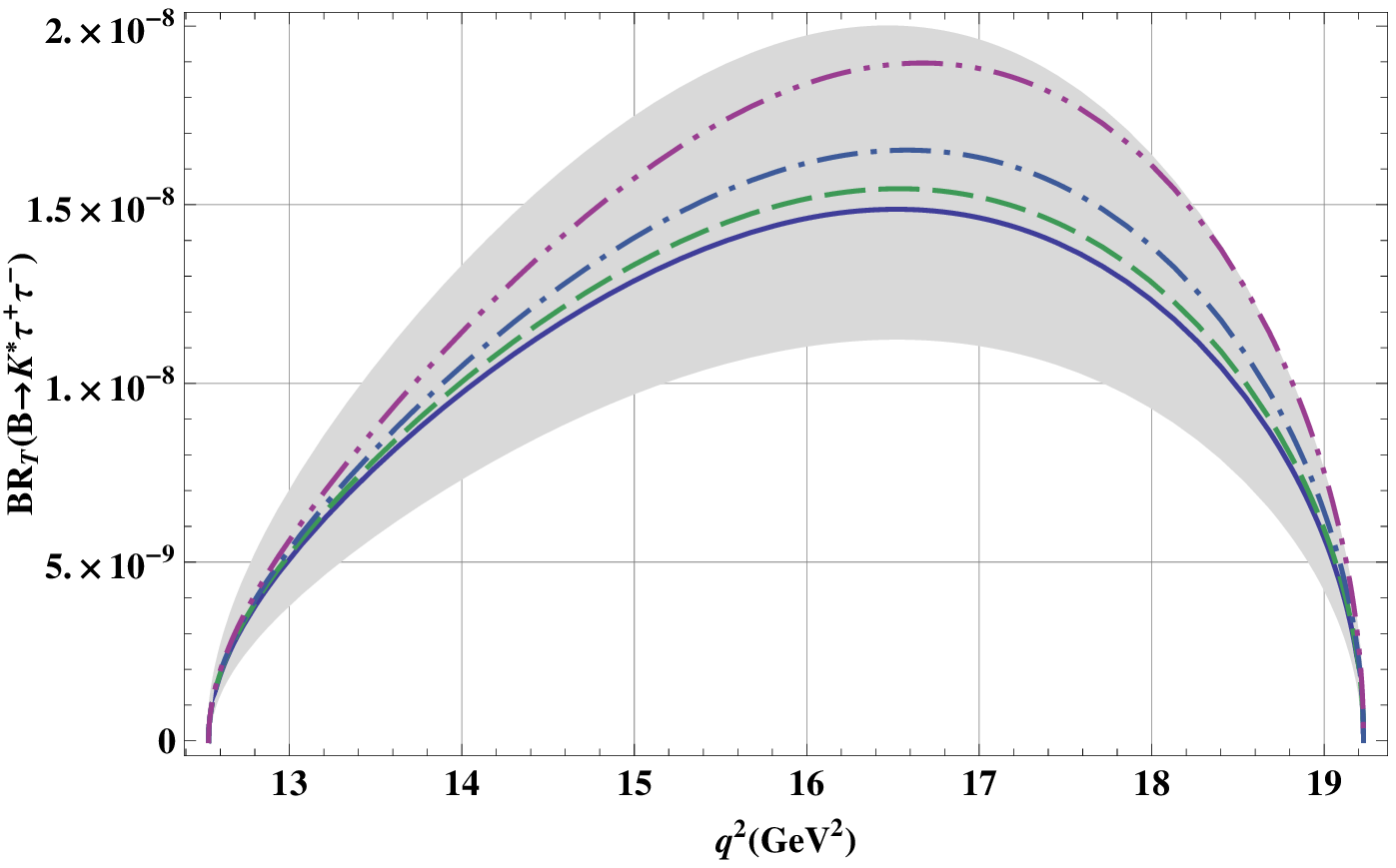} \ \ \
& \ \ \ \includegraphics[scale=0.5]{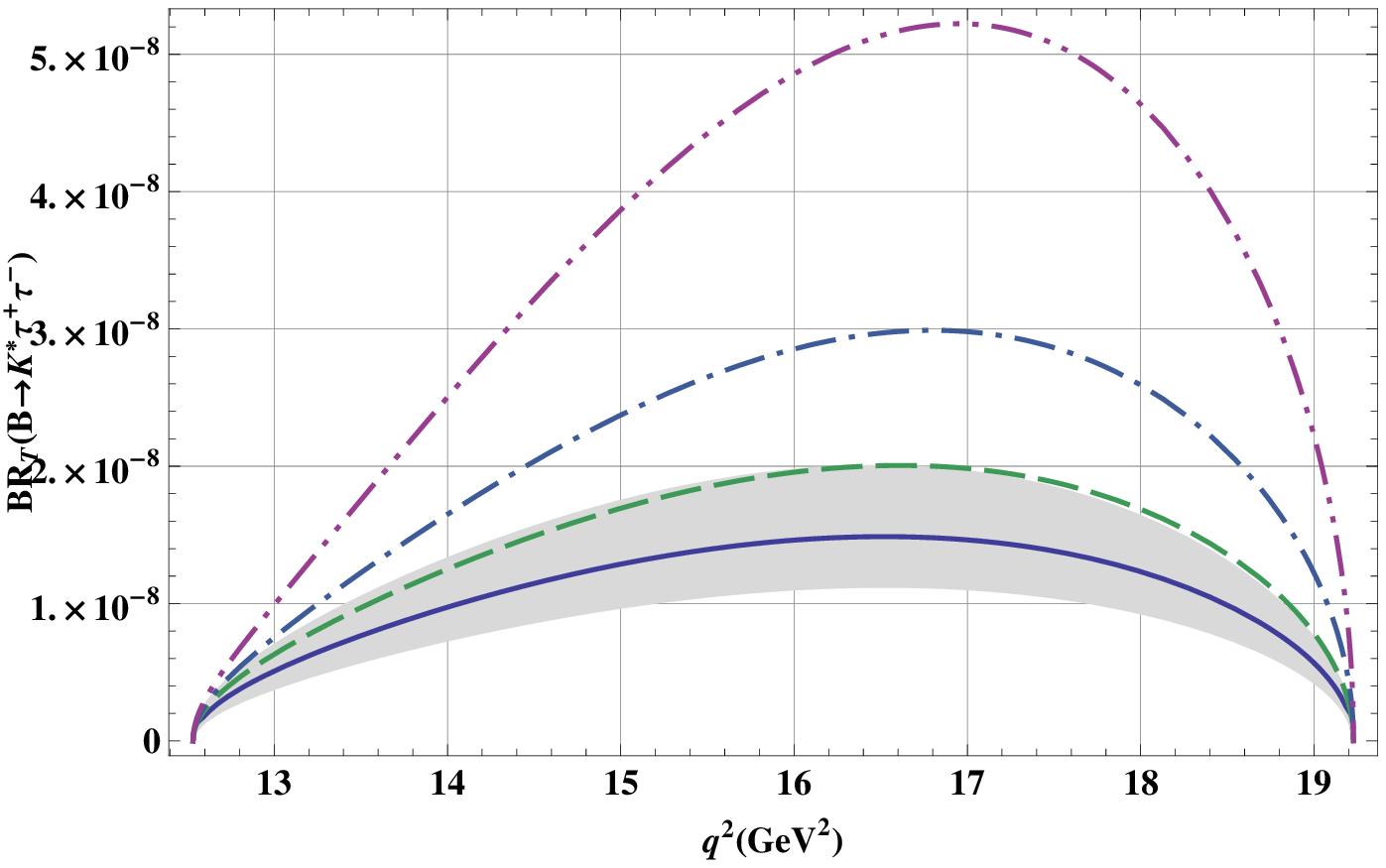}\end{tabular}
\caption{The dependence of the longitudinal and transverse BR for
the decay $B\to K^*(892)\tau^{+}\tau^{-}$ on $q^{2}$ for different
values of $m_{t^{\prime}}$ and $\left\vert
V^{\ast}_{t^{\prime}b}V_{t^{\prime}s}\right\vert$. Legends and the
values of the fourth-generation parameters are the same as in Fig.
\ref{lpm}.} \label{lpt}
\end{figure*}

\begin{figure*}[ht]
\begin{tabular}{cc}
\hspace{0.6cm}($\mathbf{a}$)&\hspace{1.2cm}($\mathbf{b}$)\\
\includegraphics[scale=0.5]{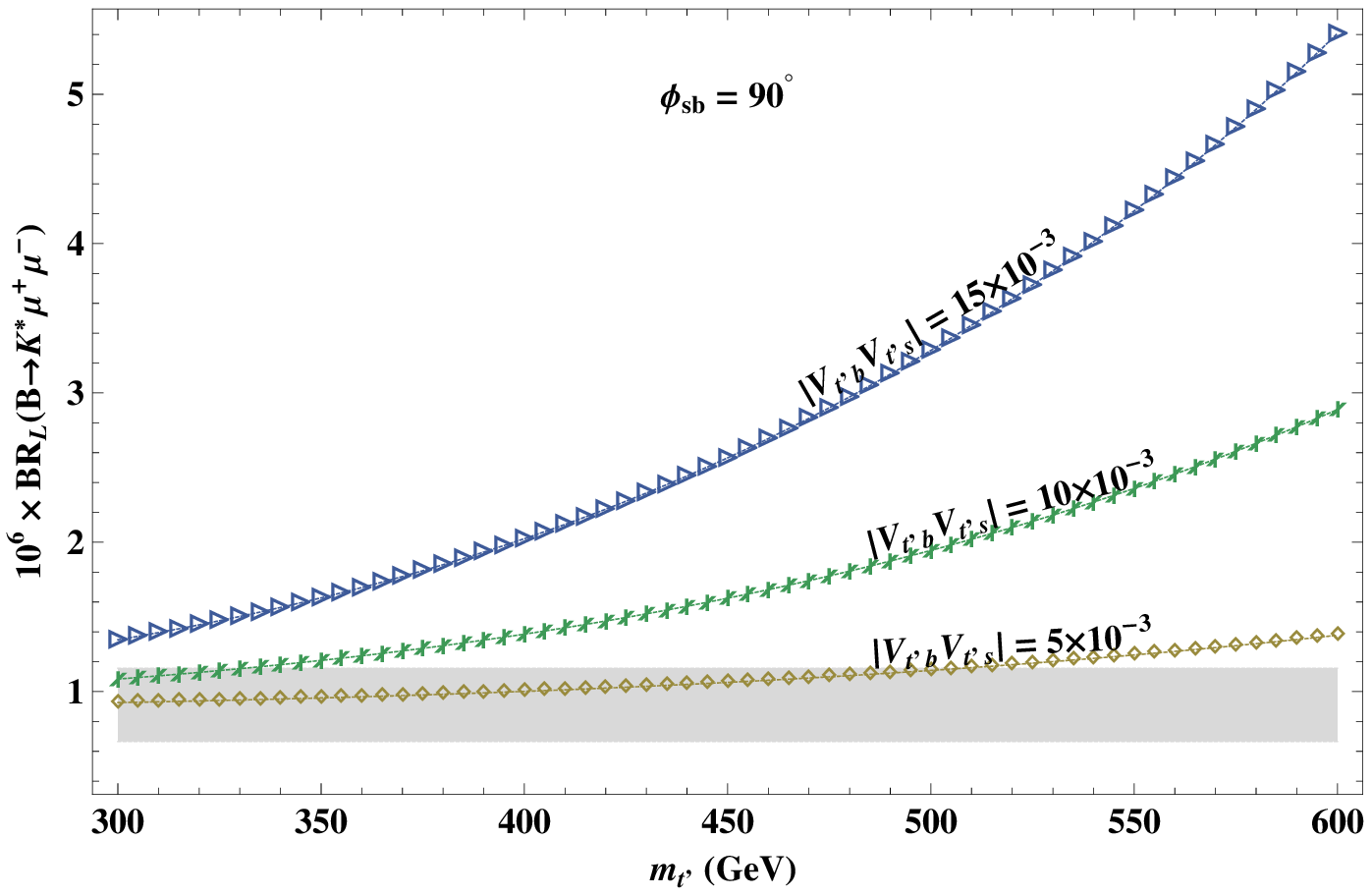} \ \ \
& \ \ \ \includegraphics[scale=0.5]{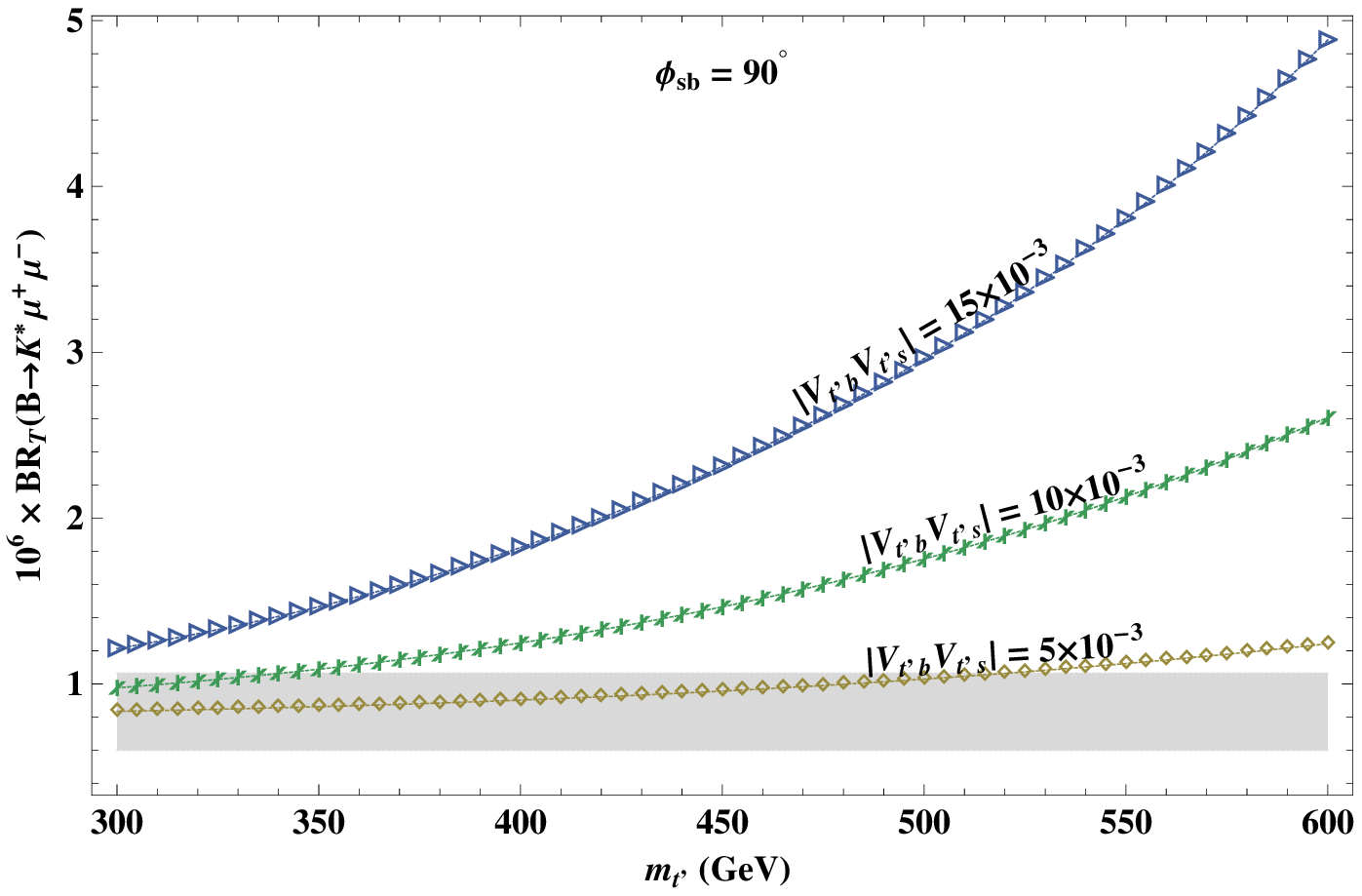}\\
\hspace{0.6cm}($\mathbf{c}$)&\hspace{1.2cm}($\mathbf{d}$)\\
\includegraphics[scale=0.5]{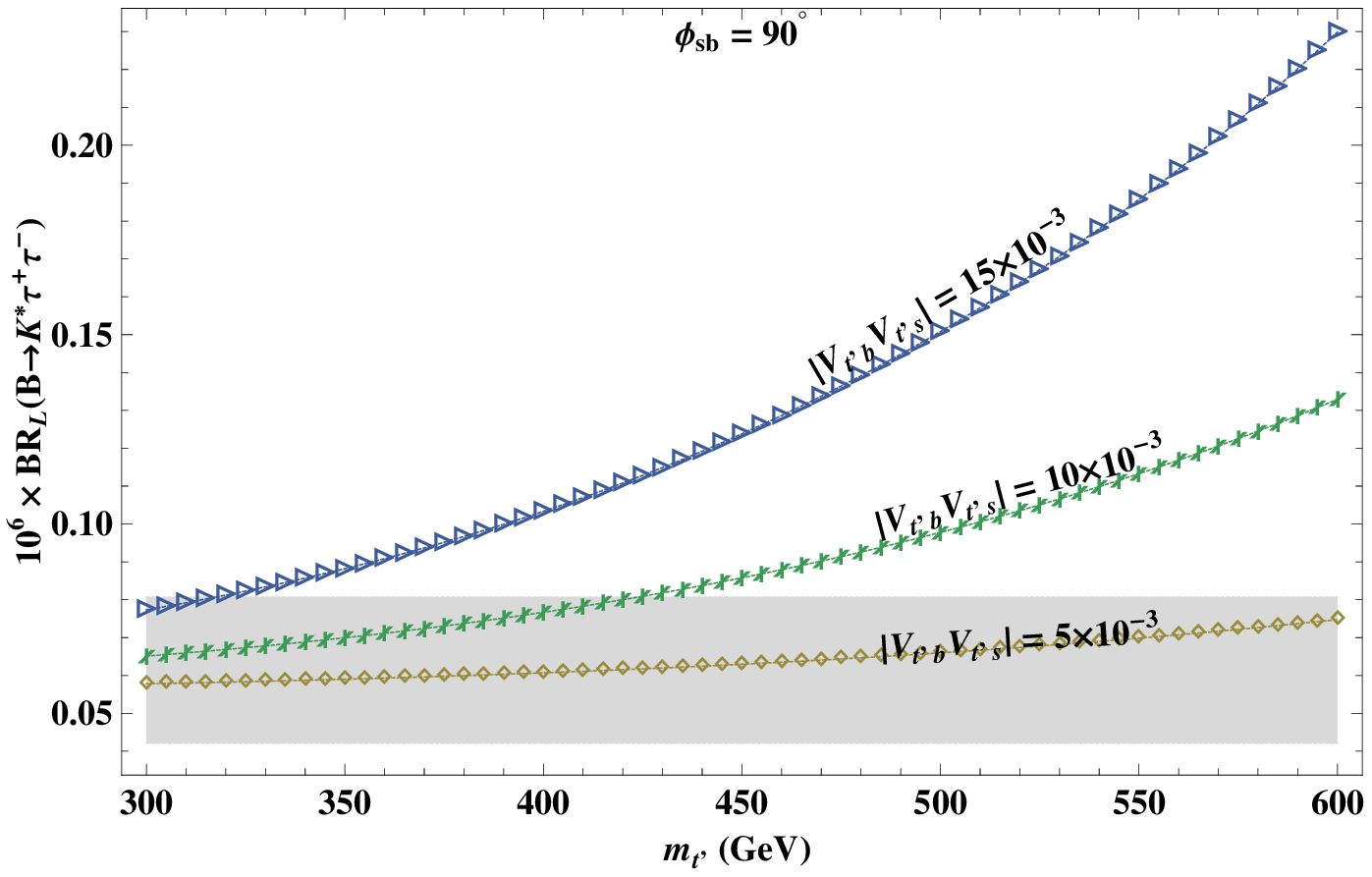} \ \ \
& \ \ \ \includegraphics[scale=0.5]{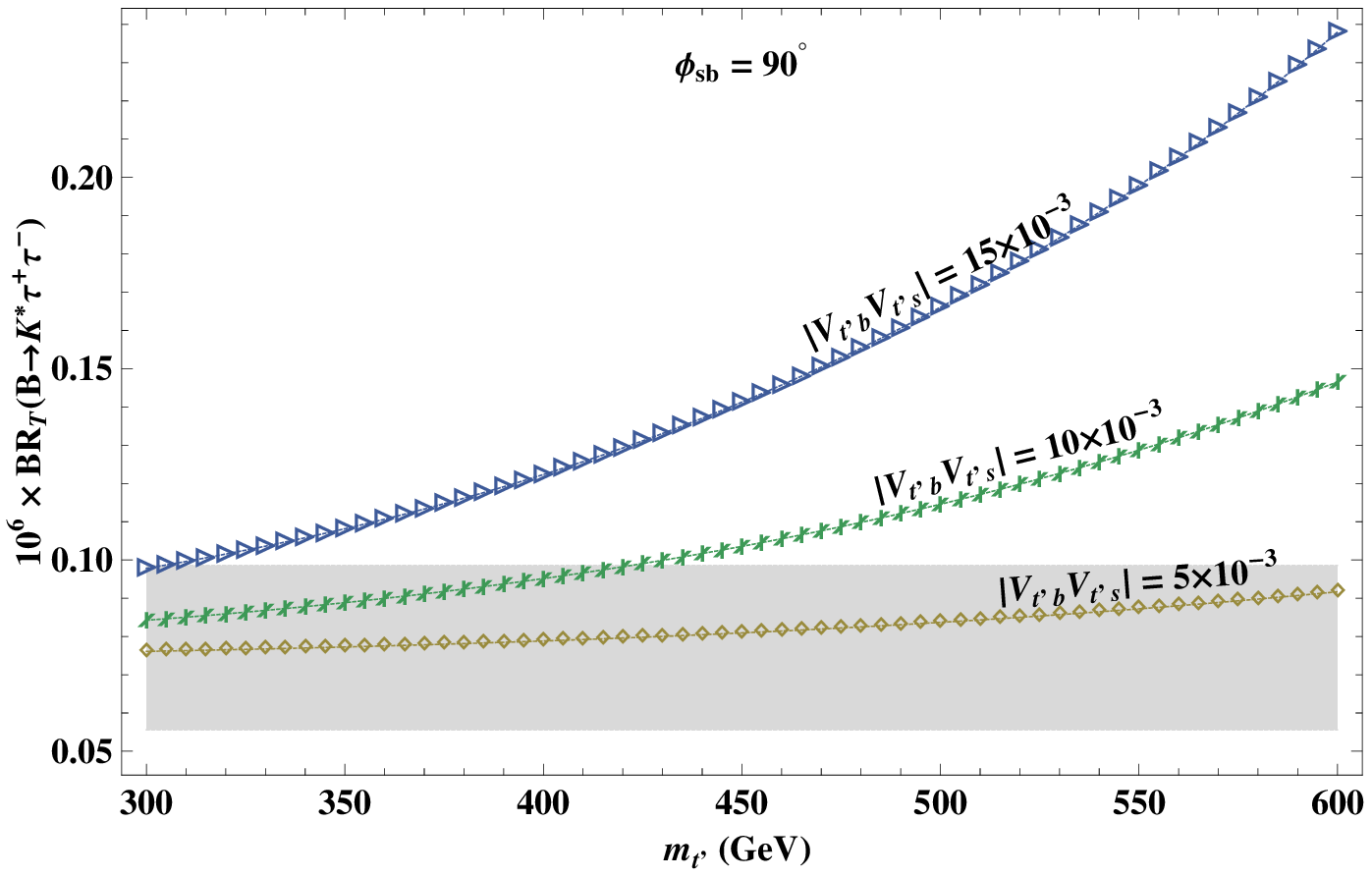}\end{tabular}
\caption{The dependence of the total longitudinal and transverse BR
for the decay $B\to K^*(892)\mu^{+}\mu^{-}$ on $m_{t'}$ for
different values of $\left\vert
V^{\ast}_{t^{\prime}b}V_{t^{\prime}s}\right\vert$.} \label{lpt}
\end{figure*}

\begin{figure*}[ht]
\begin{tabular}{cc}
\hspace{0.6cm}($\mathbf{a}$)&\hspace{1.2cm}($\mathbf{b}$)\\
\includegraphics[scale=0.5]{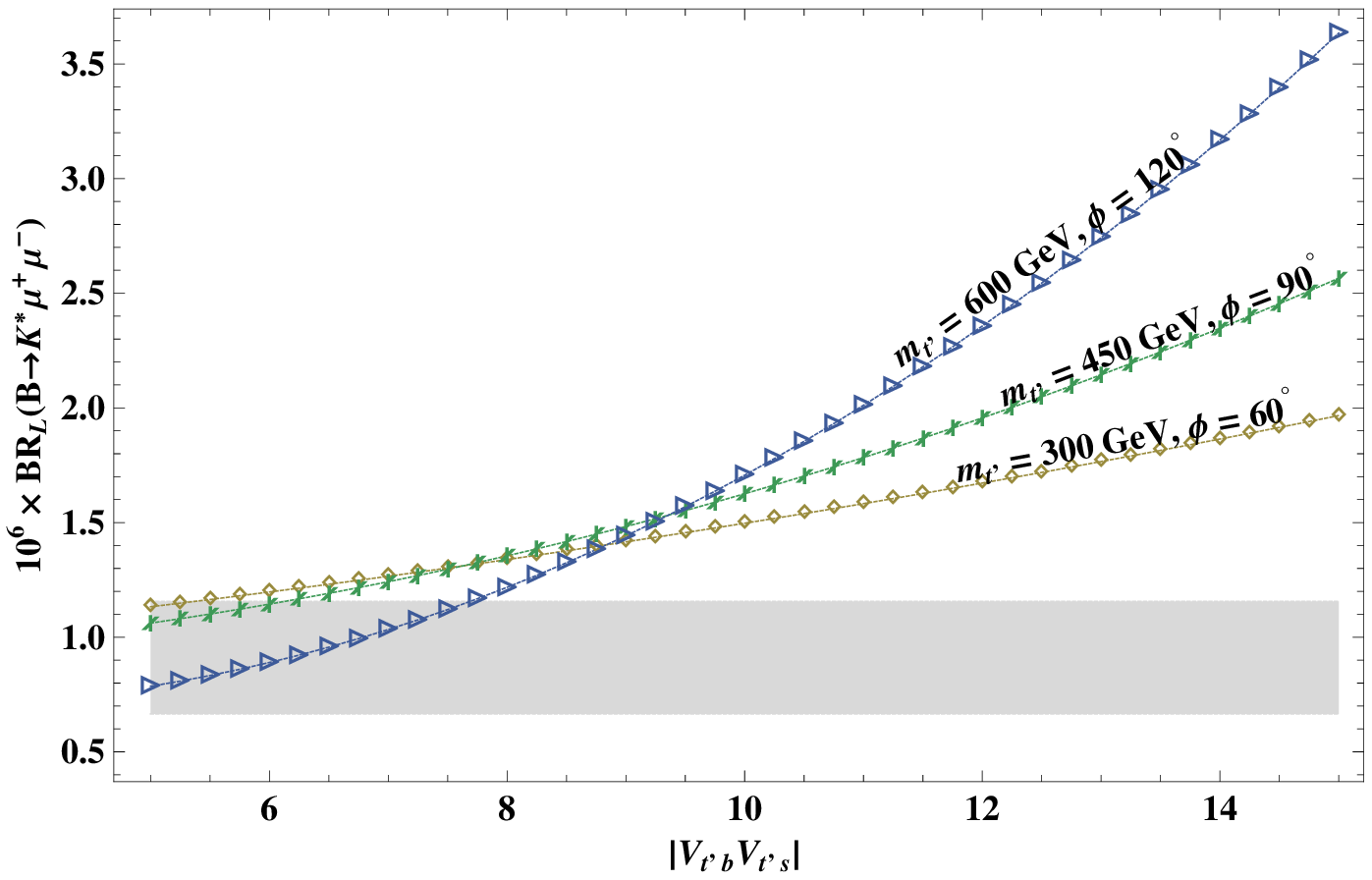} \ \ \
& \ \ \ \includegraphics[scale=0.5]{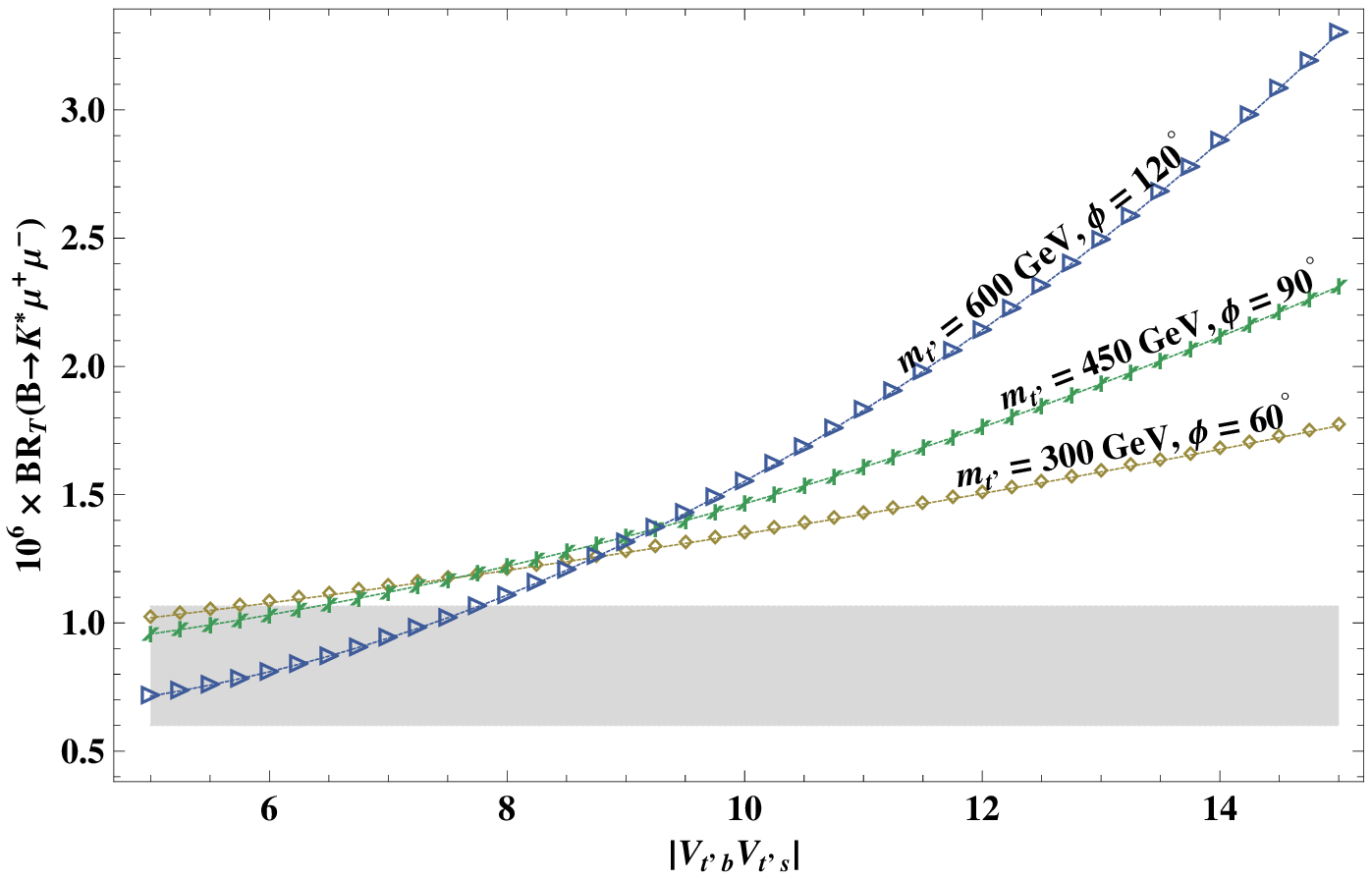}\\
\hspace{0.6cm}($\mathbf{c}$)&\hspace{1.2cm}($\mathbf{d}$)\\
\includegraphics[scale=0.5]{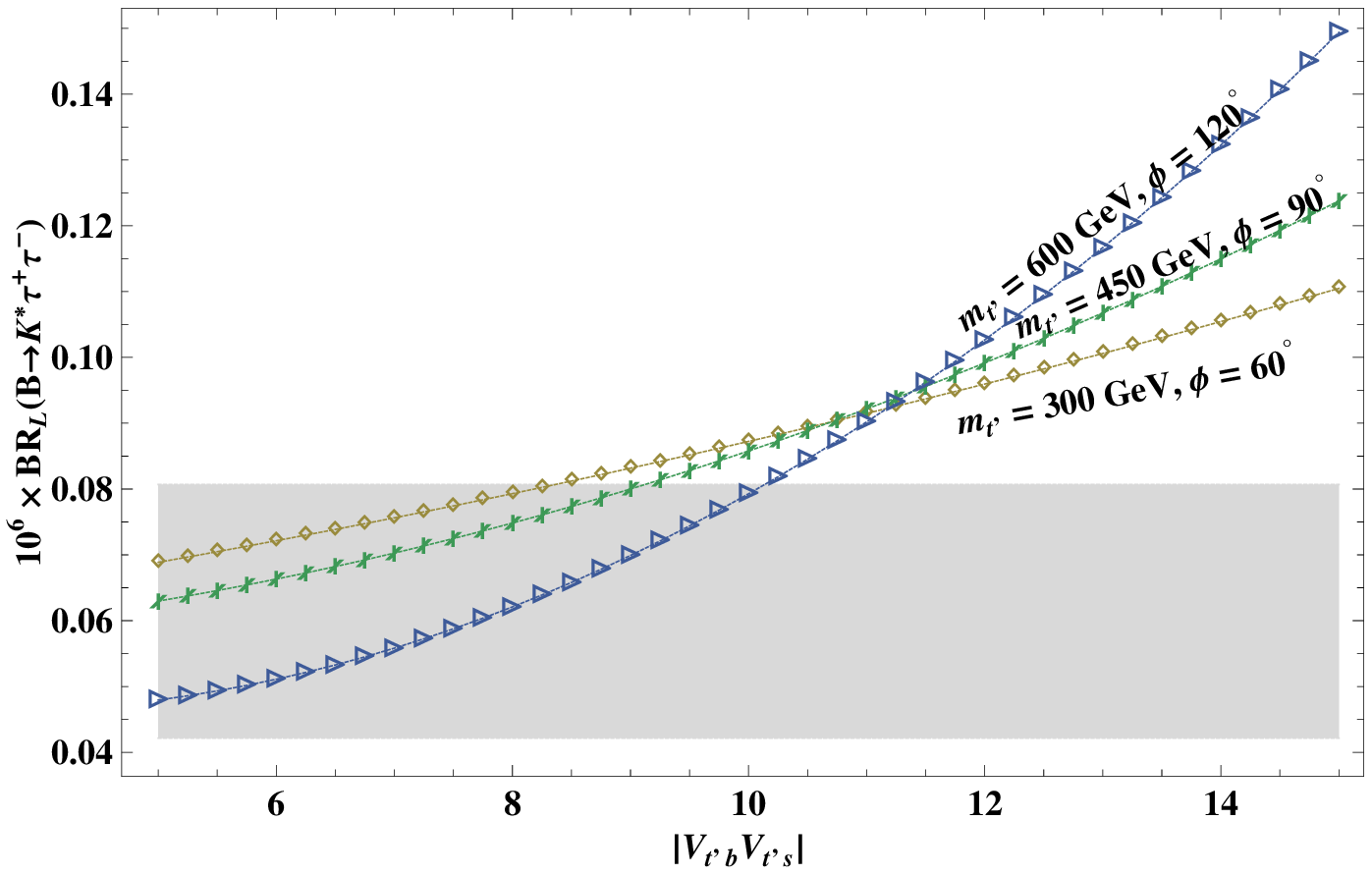} \ \ \
& \ \ \ \includegraphics[scale=0.5]{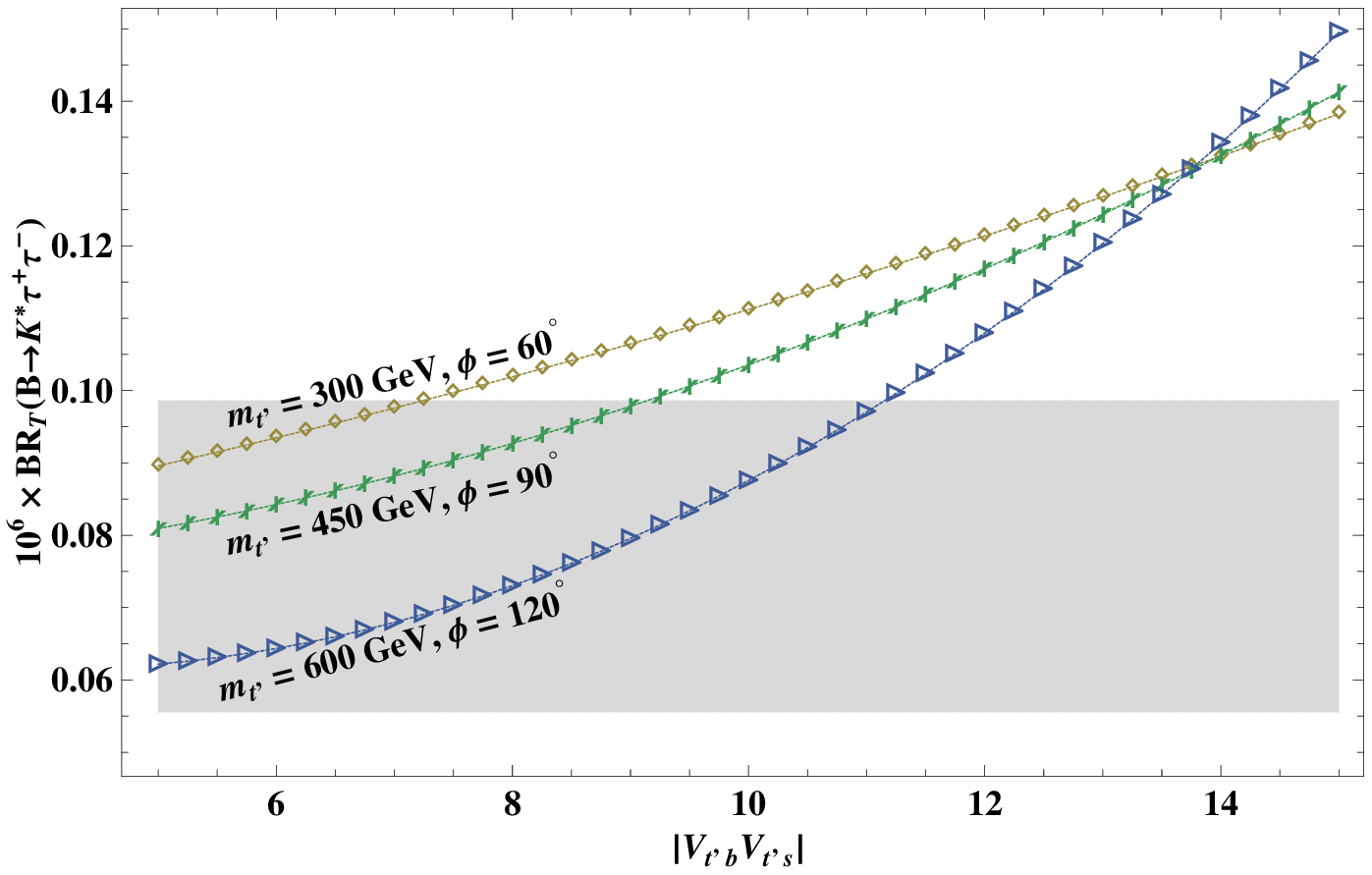}\end{tabular}
\caption{The dependence of the total longitudinal and transverse BR
for the decay $B\to K^*(892)\mu^{+}\mu^{-}$ on $\left\vert
V^{\ast}_{t^{\prime}b}V_{t^{\prime}s}\right\vert$ for different
values of $m_{t^{\prime}}$.} \label{lpt}
\end{figure*}

\begin{figure*}[ht]
\begin{tabular}{cc}
\hspace{0.6cm}($\mathbf{a}$)&\hspace{1.2cm}($\mathbf{b}$)\\
\includegraphics[scale=0.5]{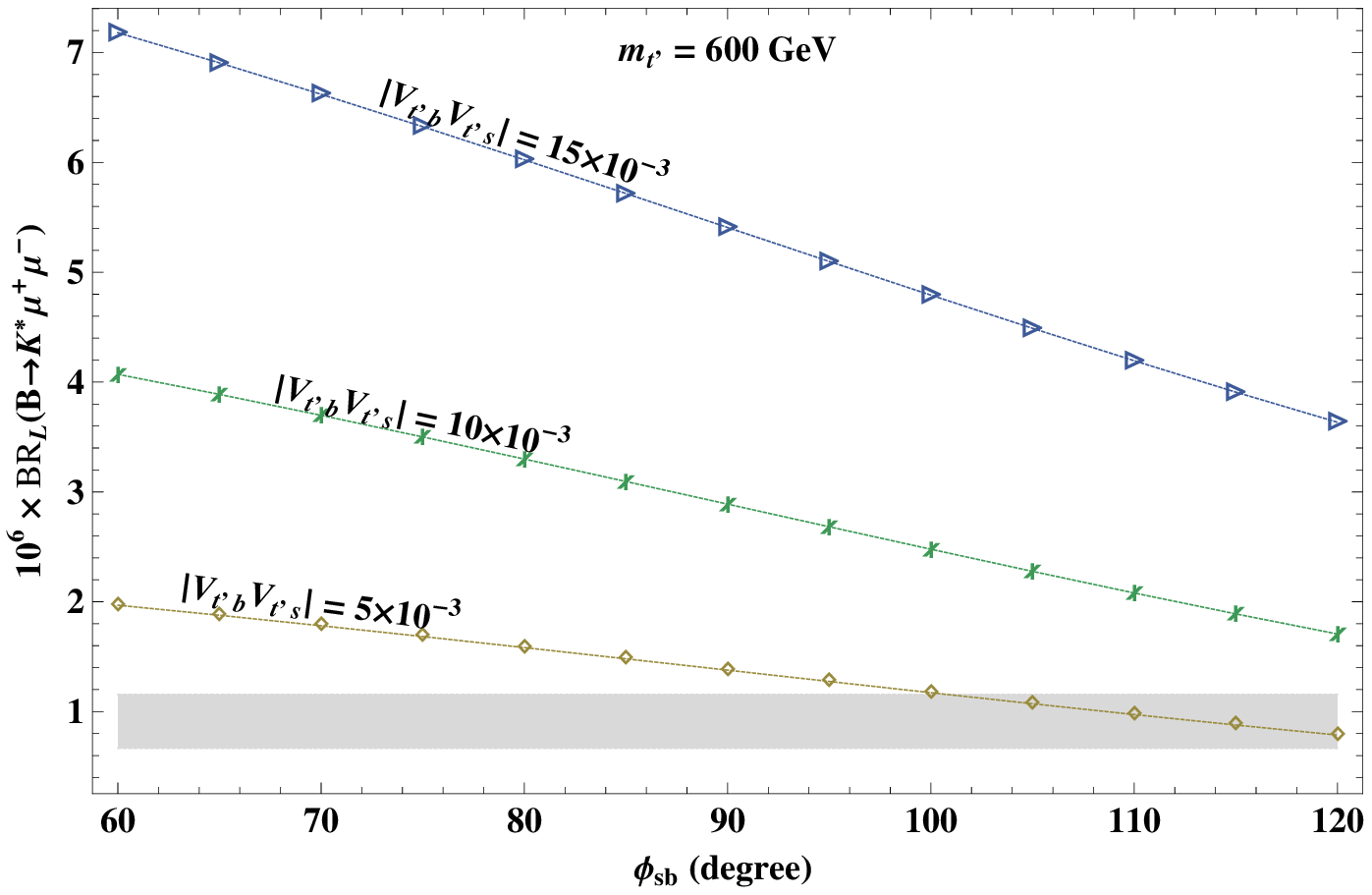} \ \ \
& \ \ \ \includegraphics[scale=0.5]{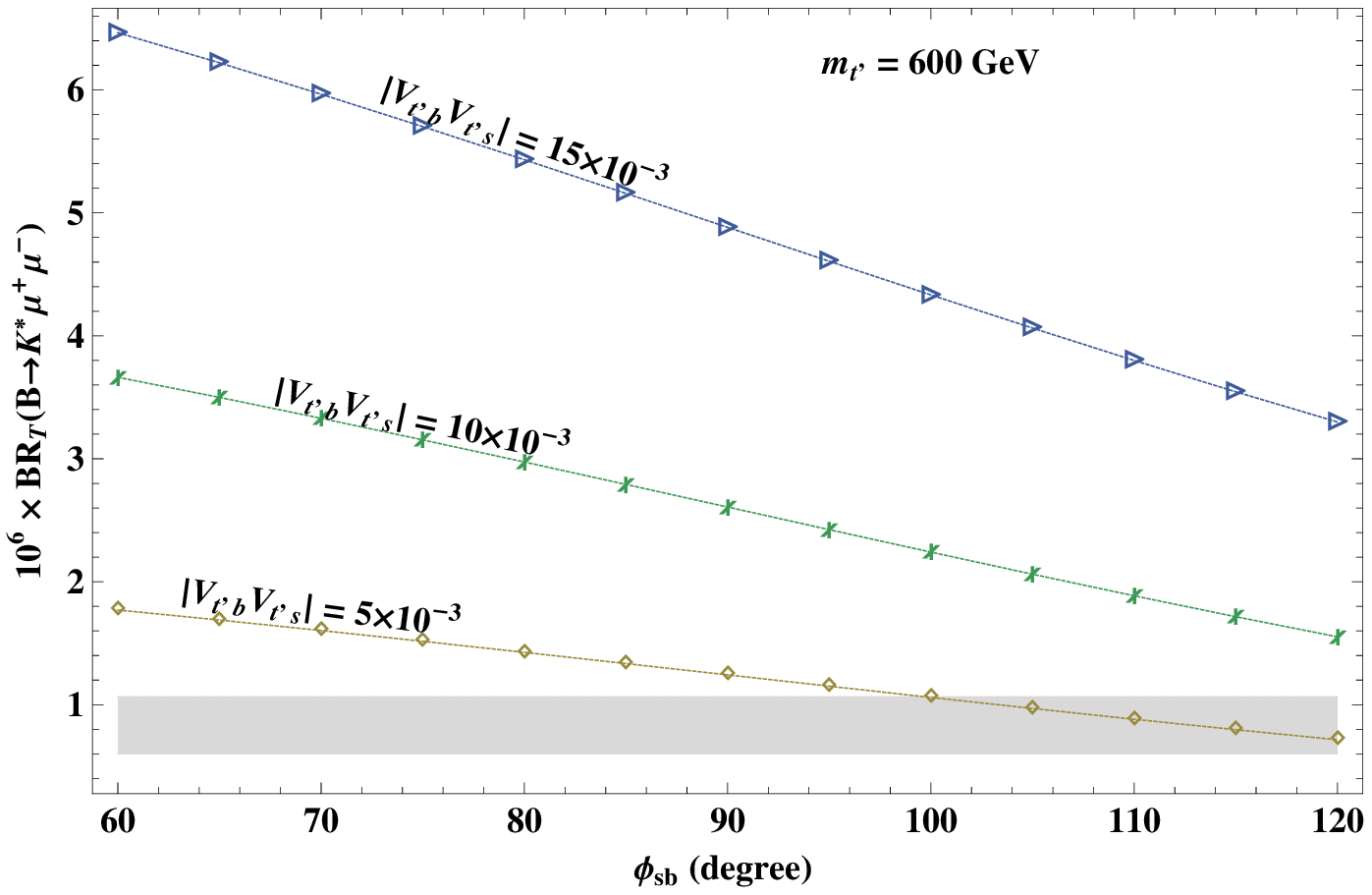}\\
\hspace{0.6cm}($\mathbf{c}$)&\hspace{1.2cm}($\mathbf{d}$)\\
\includegraphics[scale=0.5]{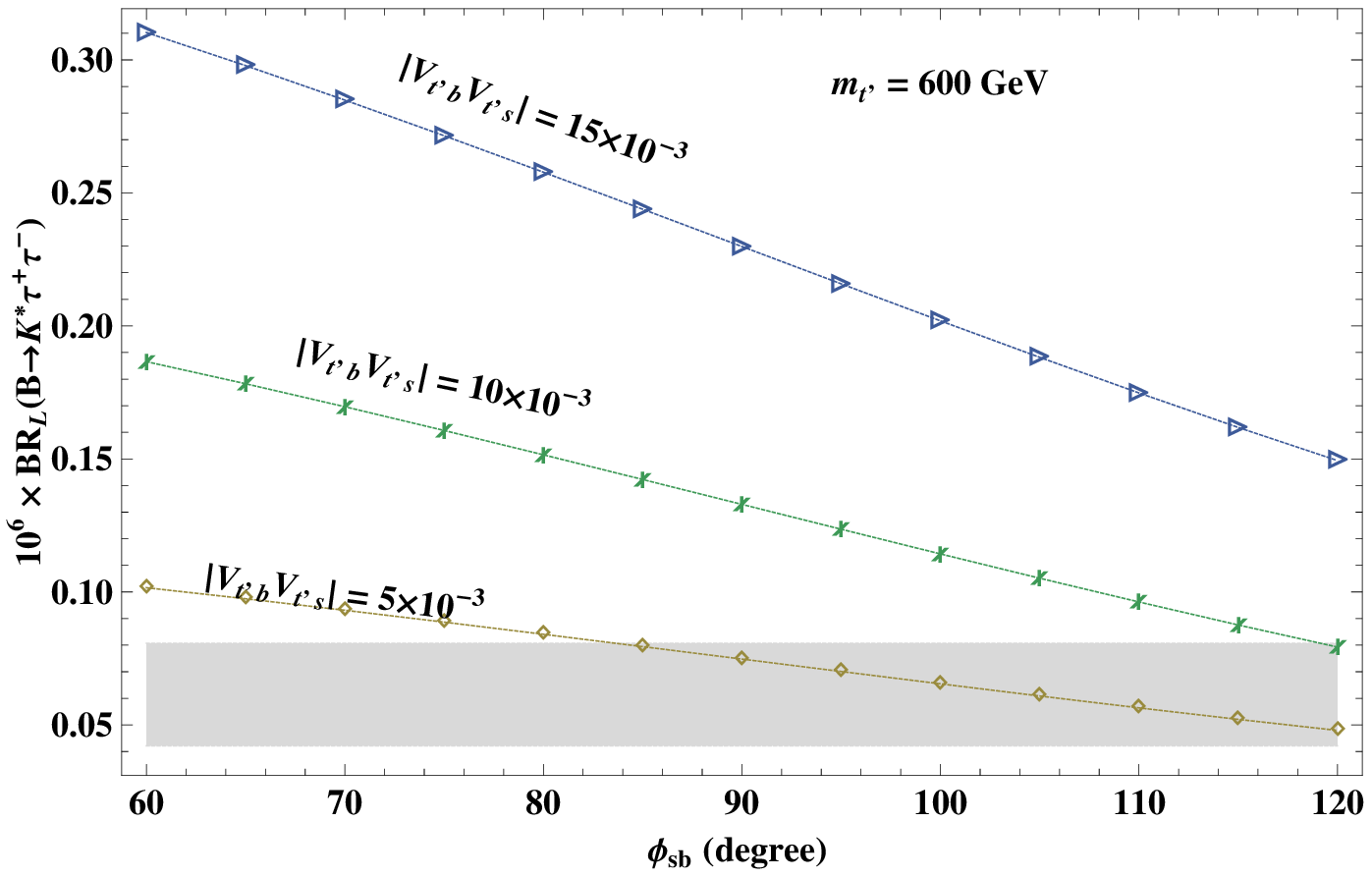} \ \ \
& \ \ \ \includegraphics[scale=0.5]{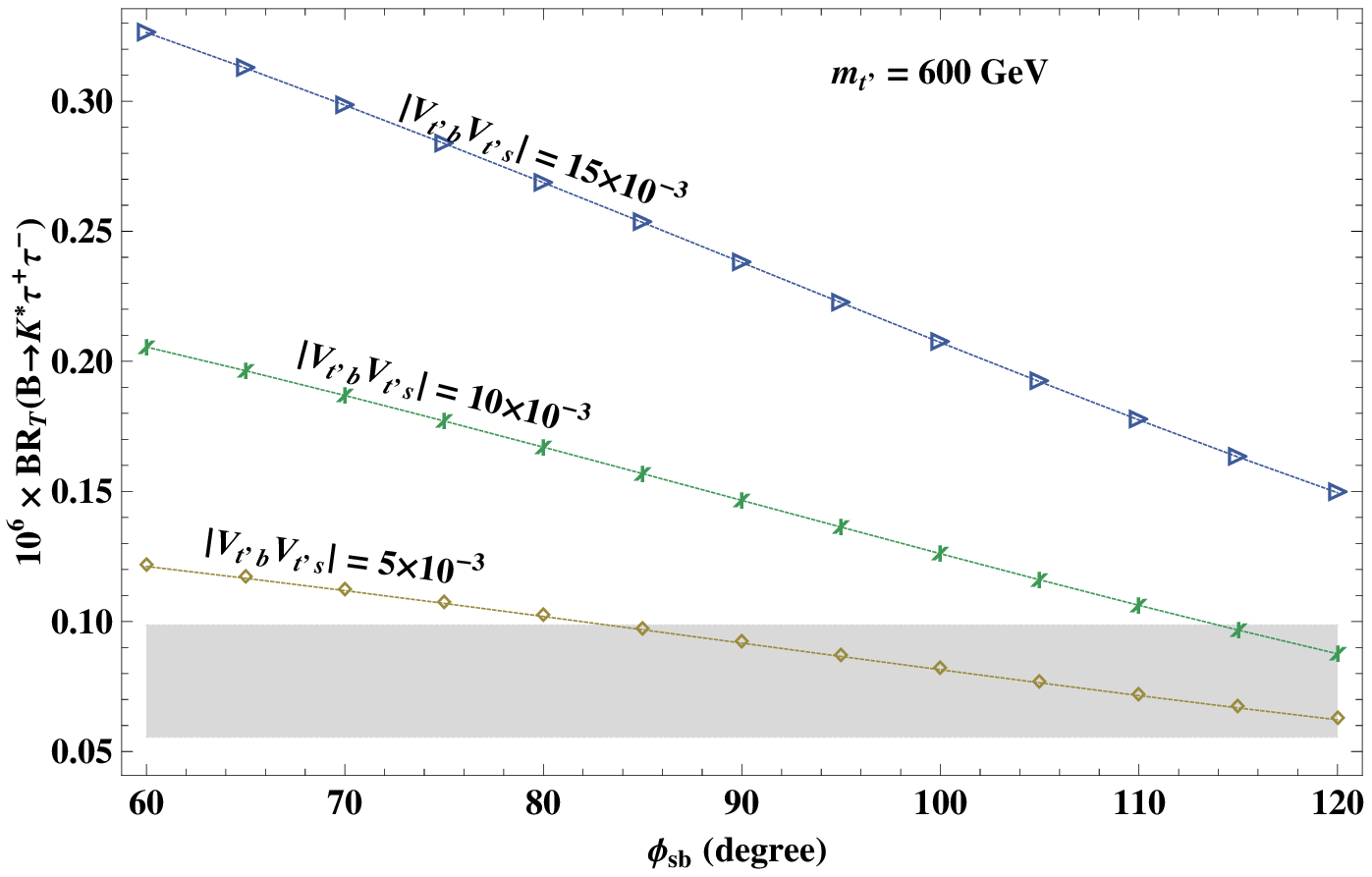}\end{tabular}
\caption{The dependence of the total longitudinal and transverse BR
for the decay $B\to K^*(892)\mu^{+}\mu^{-}$ on $\phi _{sb}$ for
different values of $m_{t^{\prime}}$ and $\left\vert
V^{\ast}_{t^{\prime}b}V_{t^{\prime}s}\right\vert$.} \label{lpt}
\end{figure*}

\subsubsection{NP in helicity fractions $f_{L}$ and $f_{T}$}

\begin{itemize}
\item As for as the study of the polarization of the final state meson $K^{*}$ is
concerned in the $B\rightarrow K^{*}l^{+}l^{-}$ decay channel the
longitudinal $f_{L}$ and transverse $f_{T}$ helicity fractions
become important observable since the uncertainty in this observable
is almost negligible, especially when we have muons as the final
state leptons. The helicity fraction is the probability of
longitudinally and transversely polarized $K^{*}$ meson in the above
mentioned decay channel so their sum should be equal to one which
can be seen in Figs. 6 and 7. Moreover, in Fig. 6(a,b) we have
punched the data points $\bullet$ (black), $\blacksquare$ (red),
$\triangledown$ (green) and $\oplus$ (orange) corresponding to the
LHCb \cite{LHCb}, CDF \cite{CDF}, Belle \cite{Belle} and Babar
\cite{babar} collaborations, respectively.
\item In Fig 6. $f_{L}$ and $f_{T}$ as a function of $q^{2}
(GeV^{2})$ for muons as final state leptons are plotted. Here, to
check the influence of SM4 on the $f_{L}$ and $f_{T}$ we set
$\phi_{sb}=90^{\circ}$ and vary the values of $m_{t'}$ and
$|V_{t'b}V_{t's}|$. Plots 6a and 6b depict that at the minimum value
of $m_{t'}=300$ GeV the effects in helicity fraction are negligible
while at the maximum value of $m_{t'}=600$ GeV the effects are mild.
Where as the recent results especially from the LHCb and CDF
collaboration is favoring the SM predictions as shown in Fig. 6(a,b)
and one can also see that the corresponding SM4 curves are very
close to some the data points and the deviation to the SM curve is
also not robust so the helicity fraction for the case of muons is
not a good observable to pin down the status of SM4. Moreover, we
have also made the estimate of the average longitudinal helicity
fraction of $K^{\ast}$ meson in the low $q^{2}$ bin ($0.1\leq q^{2}
\leq 6 \text{ GeV}^2$) in the SM4 scenario as summarized in the
Table \ref{sm4l}. One can compare these average values in the low
$q^2$ bin with that of the experimental value given in Eqs.
(\ref{64a})-(\ref{64d}). With the more data available from the LHCb
we can use this information to put constraints on the SM4 parameter
space.
\begin{table}[htp]
\centering \caption{Average longitudinal helicity fraction $\langle
f_{L} \rangle$ of $K^{\ast}$ meson for different values of
$m_{t^{\prime}}$ and $|V_{t^{\prime}b}V^{\ast}_{t^{\prime}s}|$,
whereas $\langle f_{L}^{SM} \rangle=0.649$.}
\begin{tabular}{|c|c|c|}
\hline\hline $m_{t^{\prime}}$ &
$|V_{t^{\prime}b}V^{\ast}_{t^{\prime}s}|=5\times10^{-3}$
&$|V_{t^{\prime}b}V^{\ast}_{t^{\prime}s}|=15\times10^{-3}$ \\ \hline
300 & 0.653& 0.675 \\
450 & 0.666& 0.715 \\
600 & 0.685& 0.740 \\
\hline\hline
\end{tabular}\label{sm4l}
\end{table}
\item In contrast to the case of muons the helicity fractions are
greatly influenced by SM4 when tauons are the final state leptons as
shown in Fig. 7. The effects of SM4 on the $f_{L}$ and $f_{T}$ are
clearly distinct from the corresponding SM values in the low $q^{2}$
region, while the NP effects decreased in the high $q^{2}$ region.
By a closer look at Figs. 7 (b,d) one can extract that at the
maximum values of $m_{t'}=600$ GeV and $|V_{t'b}V_{t's}|=0.015$ the
shift in the minimum(maximum) values of the $f_{T}(f_{L})$ is about
0.2 which lie at $q^{2}=4m^{2}_{\tau}$ and well measured at
experiments so this is a good observable to hunt the NP beyond the
standard model.
\item Now to qualitatively depict the effects of SM4 parameters on $f_{L}$ and $f_{T}$ for the decay
$B\rightarrow K^{*}\tau^{+}\tau^{-}$ we have displayed their average
values $<f_{L}>$ and $<f_{T}>$ as a function of $m_{t'}$,
$\phi_{sb}$ and $|V_{t'b}V_{t's}|$ in Figs. 8, 9 and 10
respectively. These graphs indicate that the $<f_{L}>$($<f_{T}>$) is
the increasing(decreasing) function of SM4 parameters. It is clear
from Figs. 8 (a, b) and 10 (a, b) that when we set
$\phi_{sb}=90^{\circ}, m_{t'}=600$ GeV and $|V_{t'b}V_{t's}|=0.015$
the value of $<f_{L}>$($<f_{T}>$) is enhanced(reduced) up(down) to
$13\%$ approximately. From figs. 9 (a, b) one can extract that at
$\phi_{sb}=120^{\circ}, m_{t'}=600$ GeV and $|V_{t'b}V_{t's}|=0.015$
this increment(decrement) in the $<f_{L}>$($<f_{T}>$) values is
reached up to $16\%$ to $17\%$ which is quite distinctive and will
be observed at LHCb.
\end{itemize}

\begin{figure*}[ht]
\begin{tabular}{cc}
\hspace{0.6cm}($\mathbf{a}$)&\hspace{1.2cm}($\mathbf{b}$)\\
\includegraphics[scale=0.5]{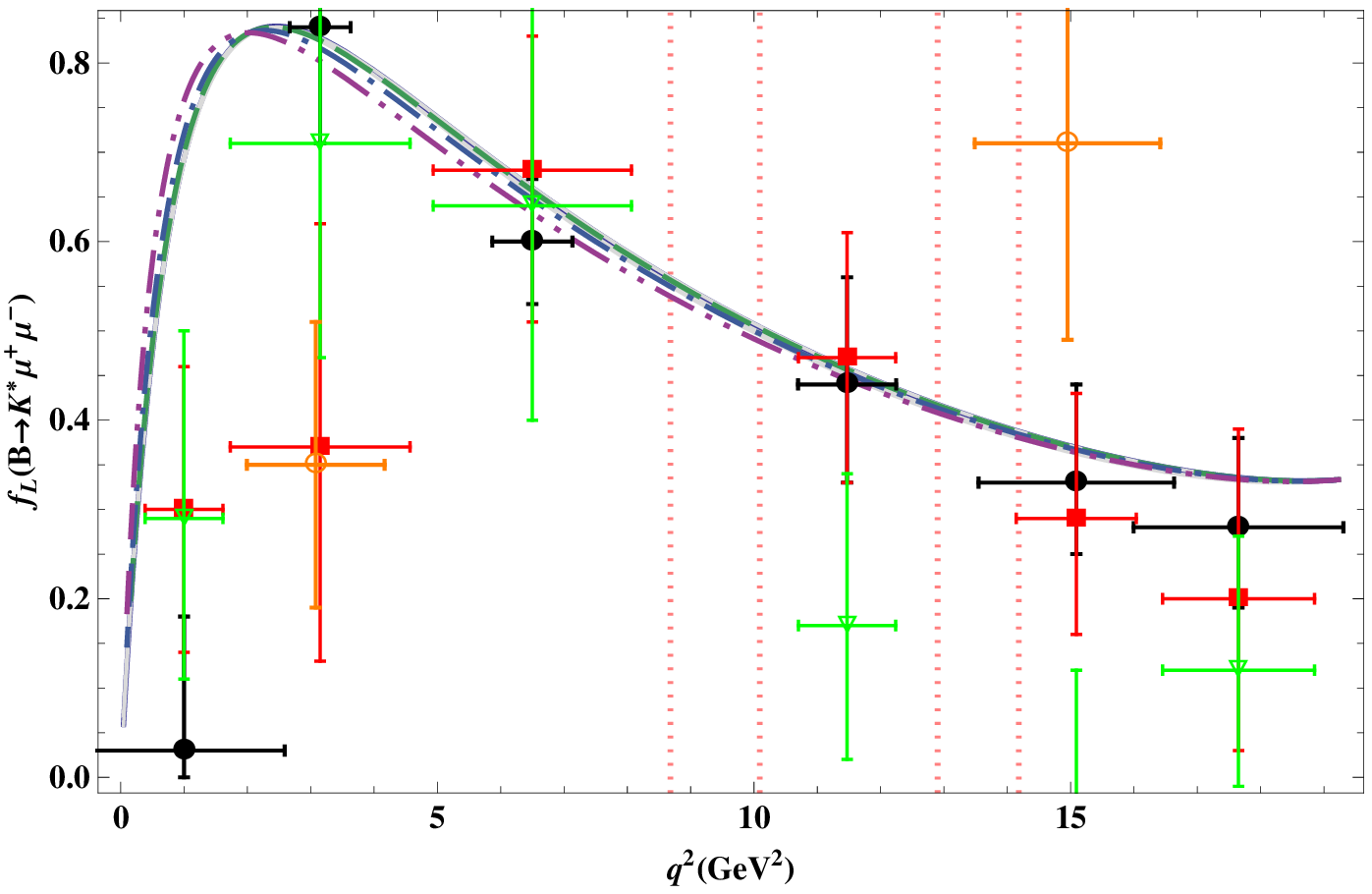} \ \ \
& \ \ \ \includegraphics[scale=0.5]{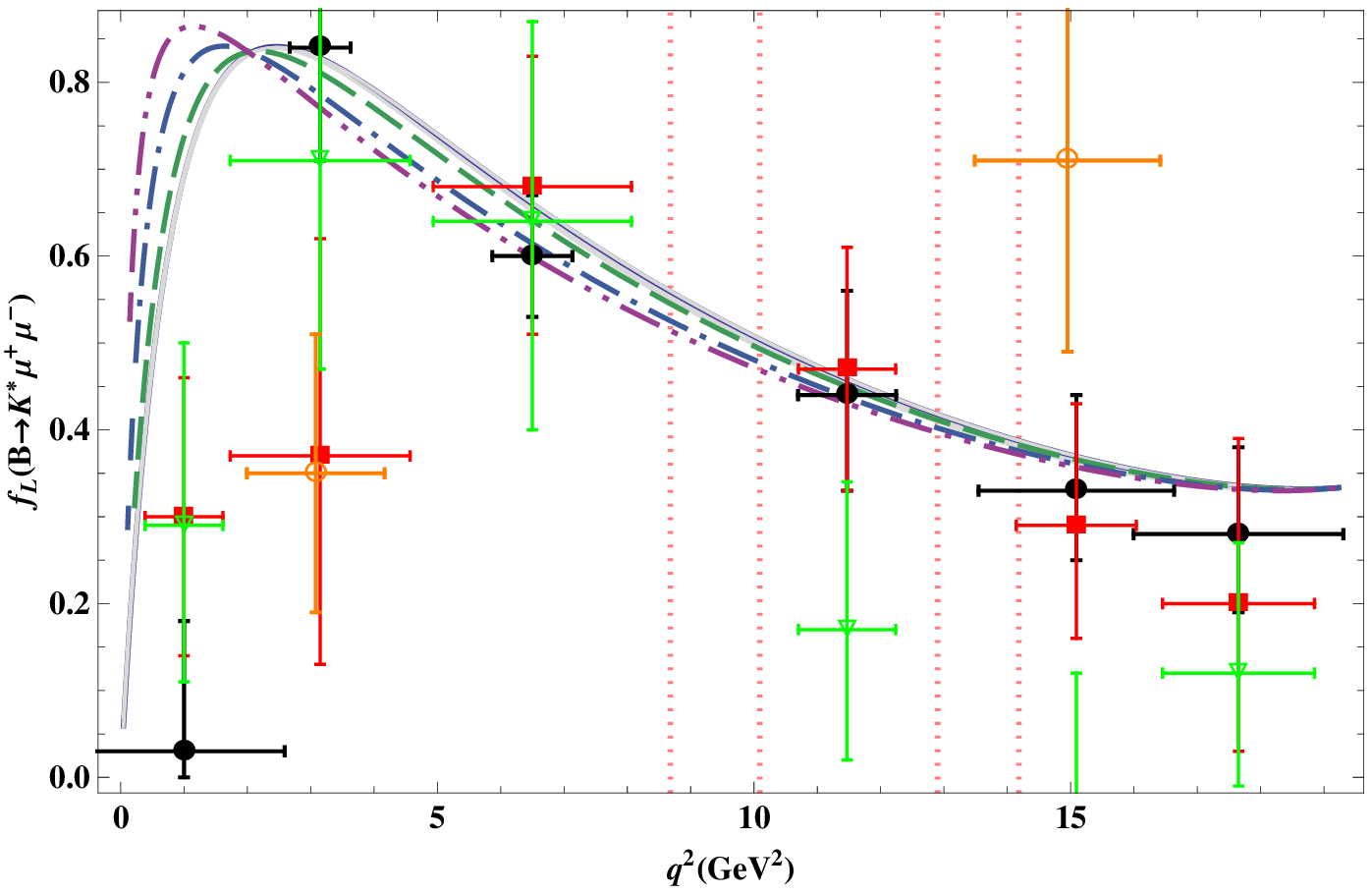}\\
\hspace{0.6cm}($\mathbf{c}$)&\hspace{1.2cm}($\mathbf{d}$)\\
\includegraphics[scale=0.5]{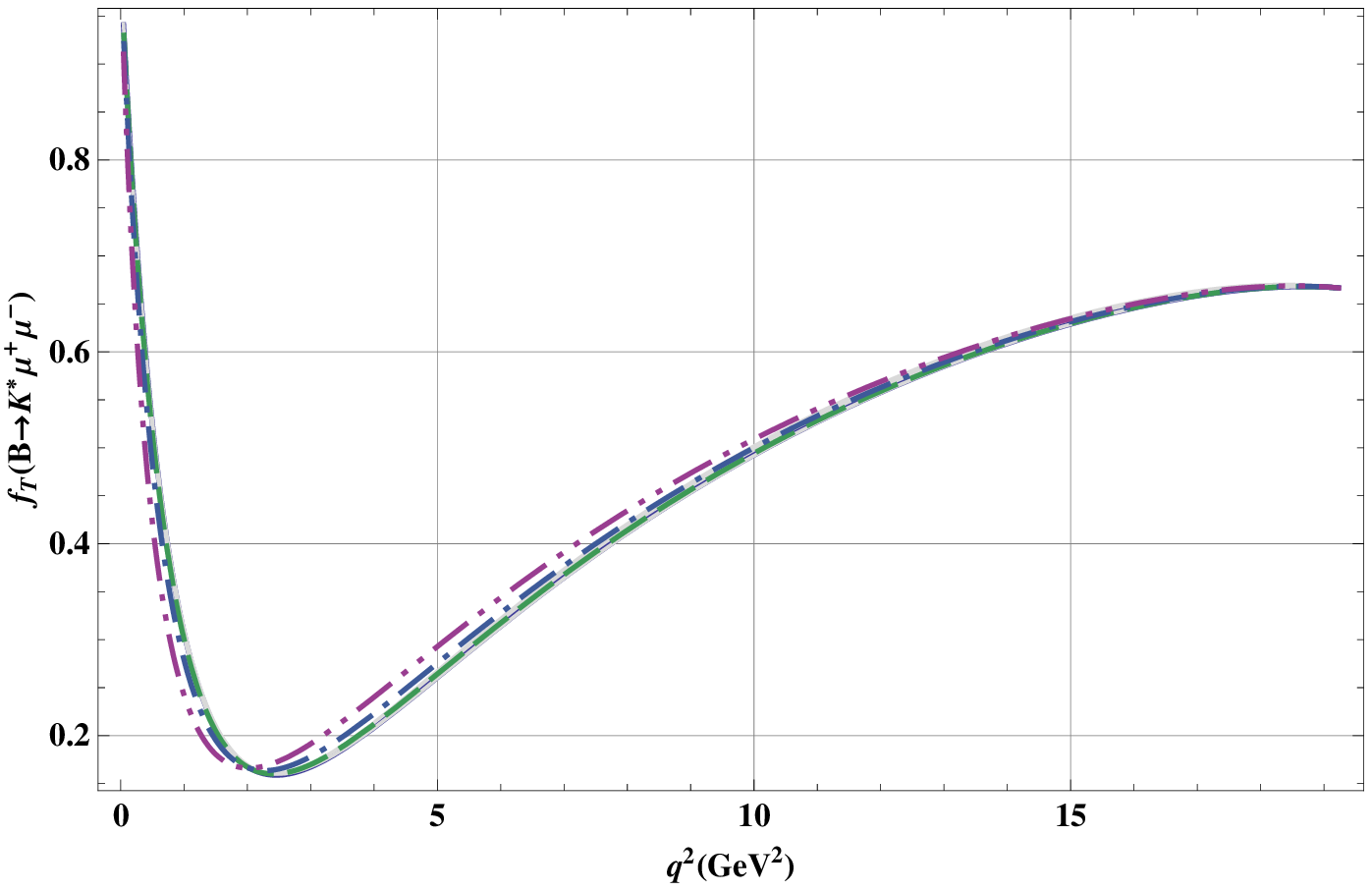} \ \ \
& \ \ \ \includegraphics[scale=0.5]{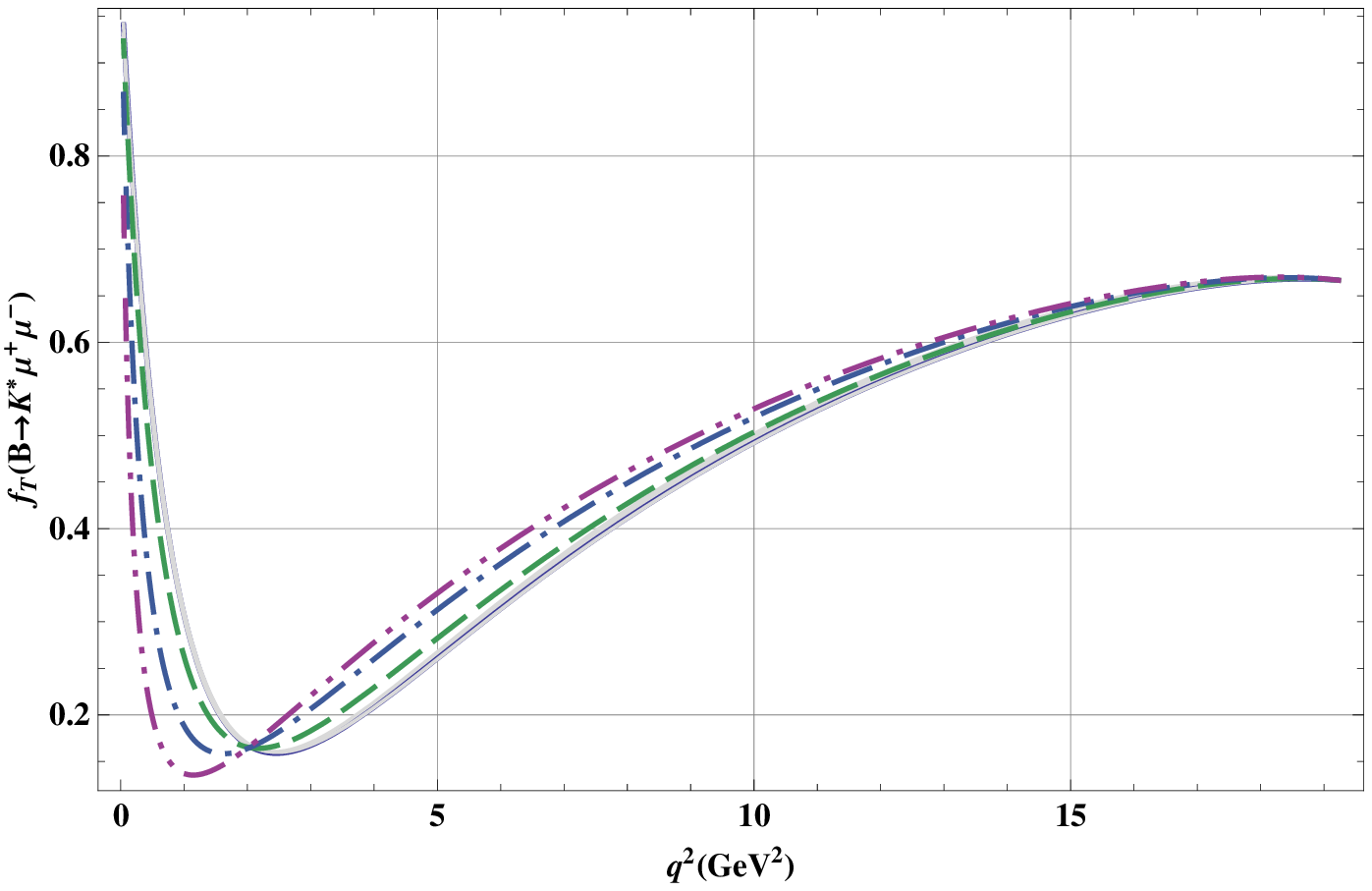}\end{tabular}
\caption{The dependence of the longitudinal and transverse helicity
fractions for the decay $B\to K^*(892)\mu^{+}\mu^{-}$ on $q^{2}$ for
different values of $m_{t^{\prime}}$ and $\left\vert
V^{\ast}_{t^{\prime}b}V_{t^{\prime}s}\right\vert$. Legends and the
values of the fourth-generation parameters are the same as in Fig.
\ref{lpm} while in the graphs in (a) and (b) the data points
$\bullet$ (black), $\blacksquare$ (red), $\triangledown$ (green) and
$\oplus$ (orange) correspond to the LHCb \cite{LHCb}, CDF
\cite{CDF}, Belle \cite{Belle} and Babar \cite{babar}
collaborations, respectively.} \label{lpt}
\end{figure*}

\begin{figure*}[ht]
\begin{tabular}{cc}
\hspace{0.6cm}($\mathbf{a}$)&\hspace{1.2cm}($\mathbf{b}$)\\
\includegraphics[scale=0.5]{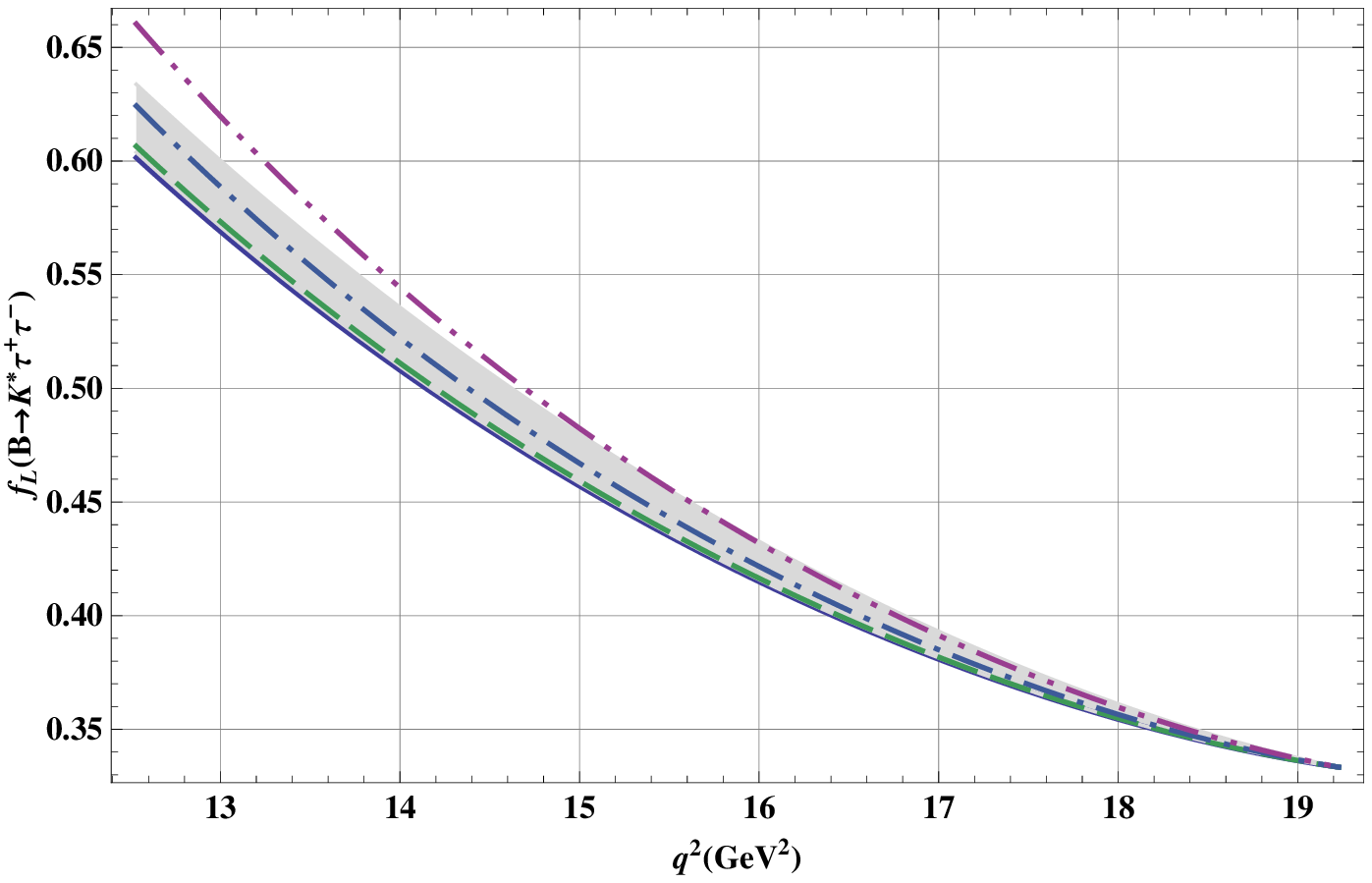} \ \ \
& \ \ \ \includegraphics[scale=0.5]{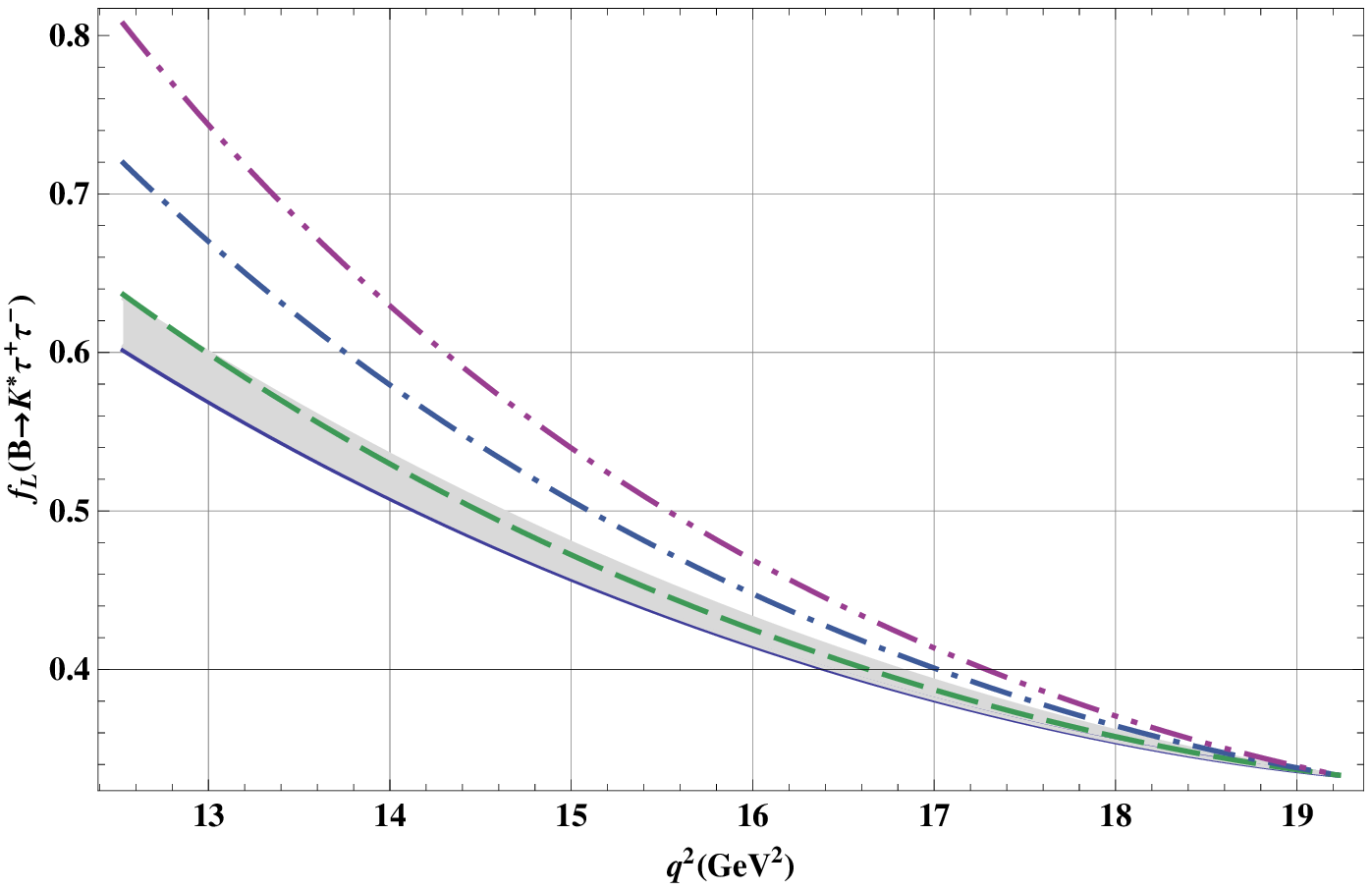}\\
\hspace{0.6cm}($\mathbf{c}$)&\hspace{1.2cm}($\mathbf{d}$)\\
\includegraphics[scale=0.5]{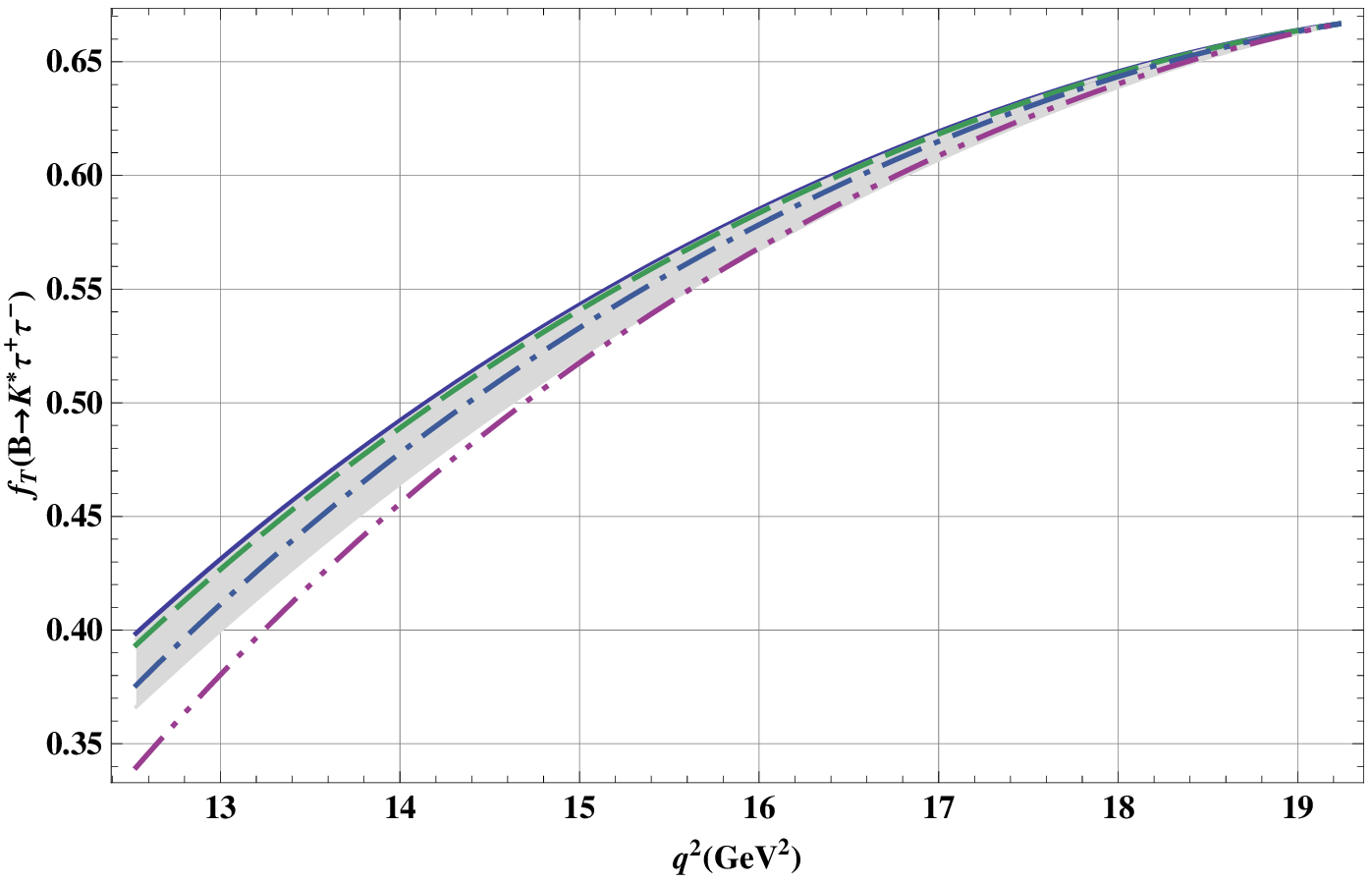} \ \ \
& \ \ \ \includegraphics[scale=0.5]{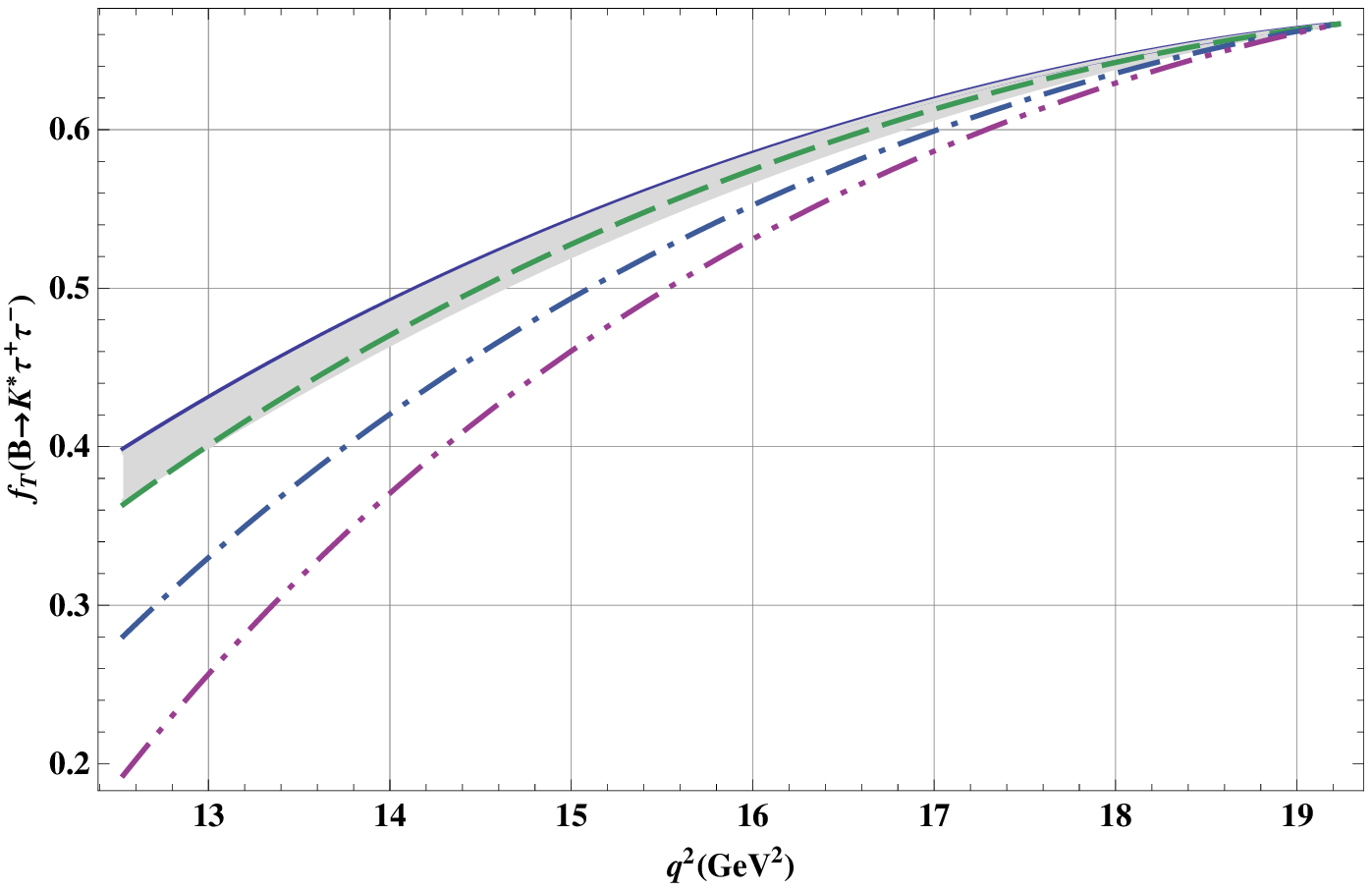}\end{tabular}
\caption{The dependence of the longitudinal and transverse helicity
fractions for the decay $B\to K^*(892)\tau^{+}\tau^{-}$ on $q^{2}$
for different values of $m_{t^{\prime}}$ and $\left\vert
V^{\ast}_{t^{\prime}b}V_{t^{\prime}s}\right\vert$. Legends and the
values of the fourth-generation parameters are the same as in Fig.
\ref{lpm}.} \label{lpt}
\end{figure*}

\begin{figure*}[ht]
\begin{tabular}{cc}
\hspace{0.6cm}($\mathbf{a}$)&\hspace{1.2cm}($\mathbf{b}$)\\
\includegraphics[scale=0.5]{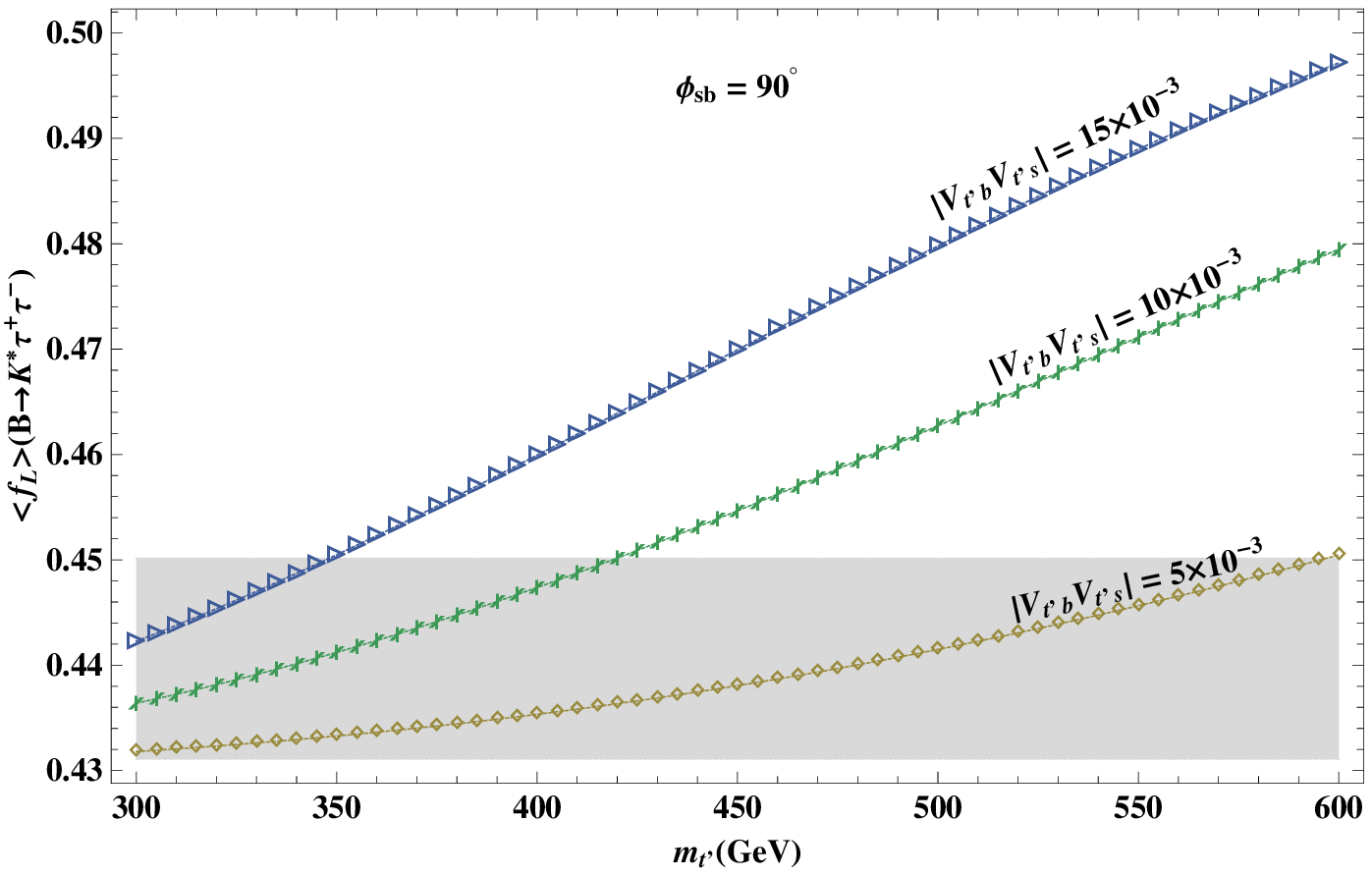} \ \ \
& \ \ \ \includegraphics[scale=0.5]{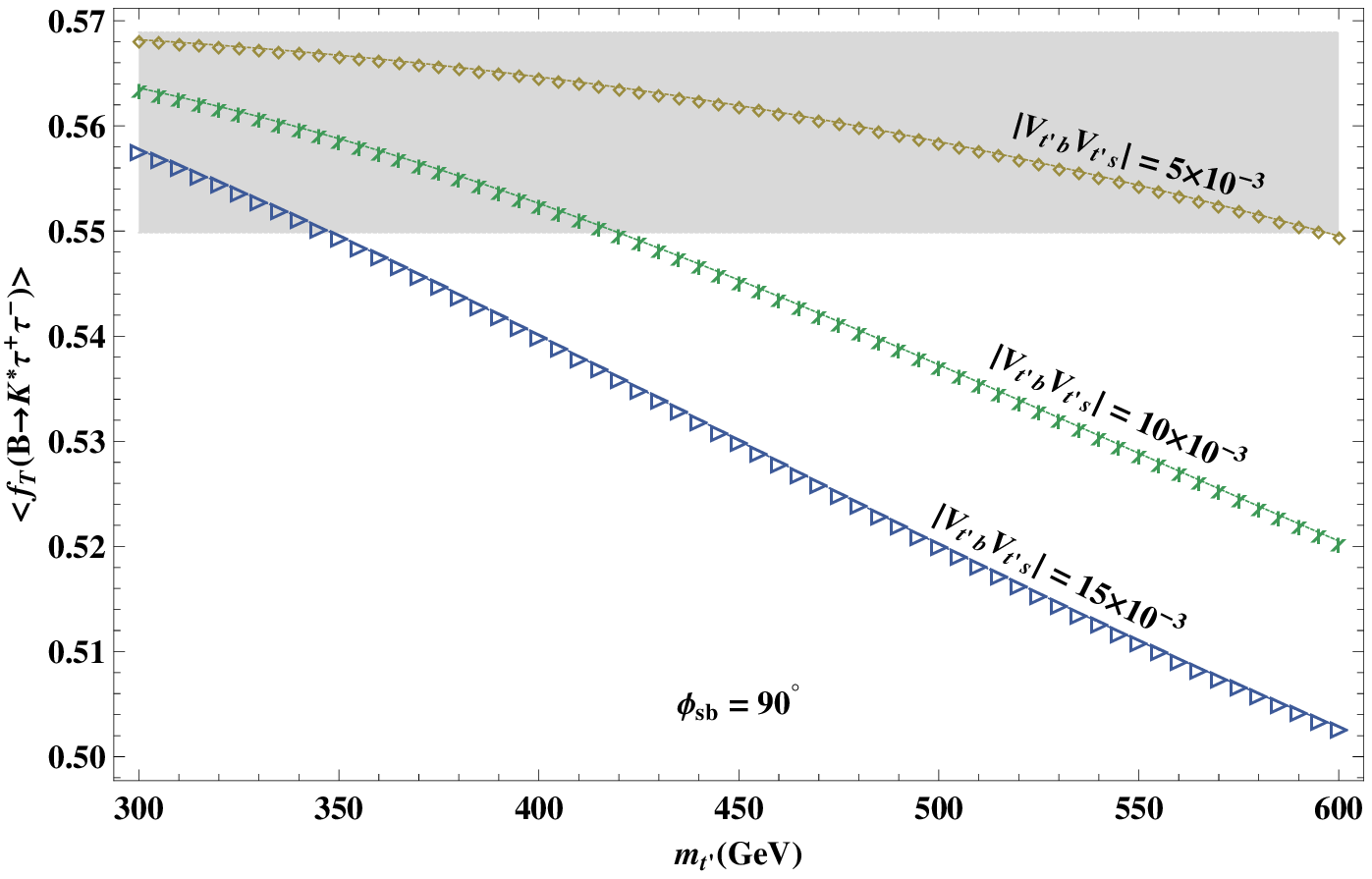}\end{tabular}
\caption{The dependence of the average longitudinal helicity
fraction for the decay $B\to K^*(892)\tau^{+}\tau^{-}$ on
$m_{t^{\prime}}$ and $\phi _{sb}$ for different values of
$\left\vert V^{\ast}_{t^{\prime}b}V_{t^{\prime}s}\right\vert$.}
\label{lpt}
\end{figure*}

\begin{figure*}[ht]
\begin{tabular}{cc}
\hspace{0.6cm}($\mathbf{a}$)&\hspace{1.2cm}($\mathbf{b}$)\\
\includegraphics[scale=0.5]{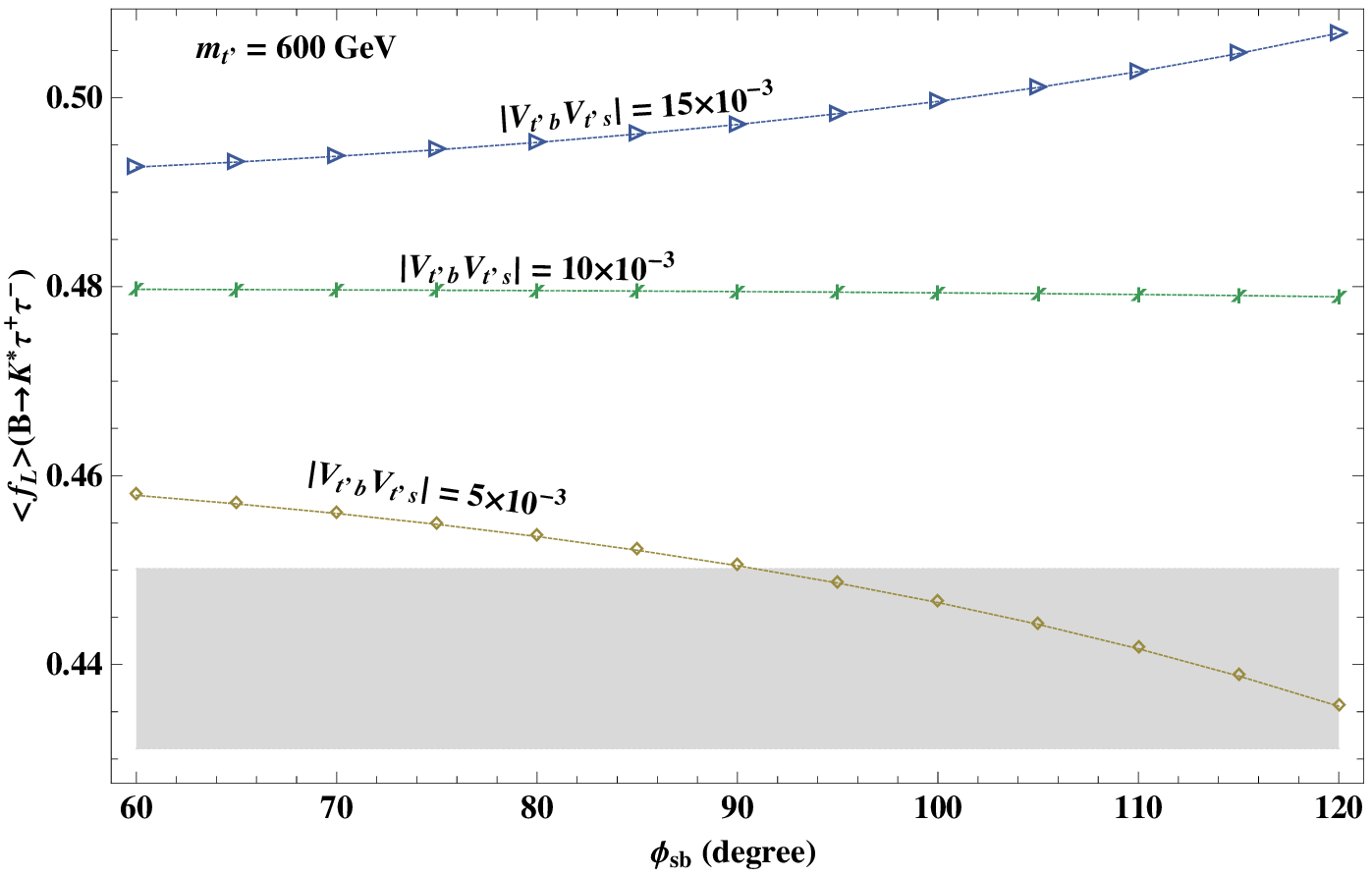} \ \ \
& \ \ \ \includegraphics[scale=0.5]{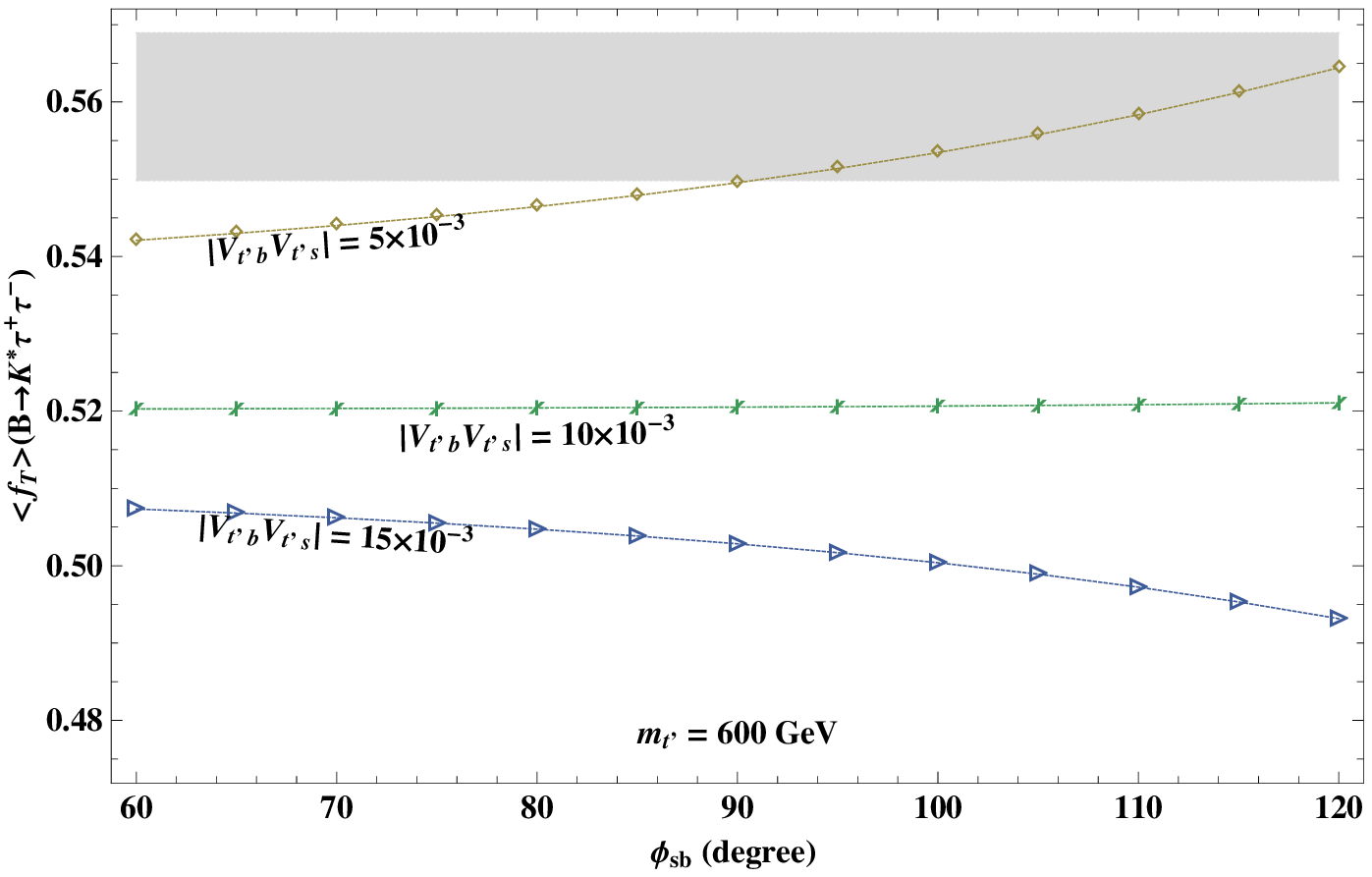}\end{tabular}
\caption{The dependence of the average transverse helicity fraction
for the decay $B\to K^*(892)\tau^{+}\tau^{-}$ on $m_{t'}$ and $\phi
_{sb}$ for different values of $\left\vert
V^{\ast}_{t^{\prime}b}V_{t^{\prime}s}\right\vert$. } \label{lpt}
\end{figure*}

\begin{figure*}[ht]
\begin{tabular}{cc}
\hspace{0.6cm}($\mathbf{a}$)&\hspace{1.2cm}($\mathbf{b}$)\\
\includegraphics[scale=0.5]{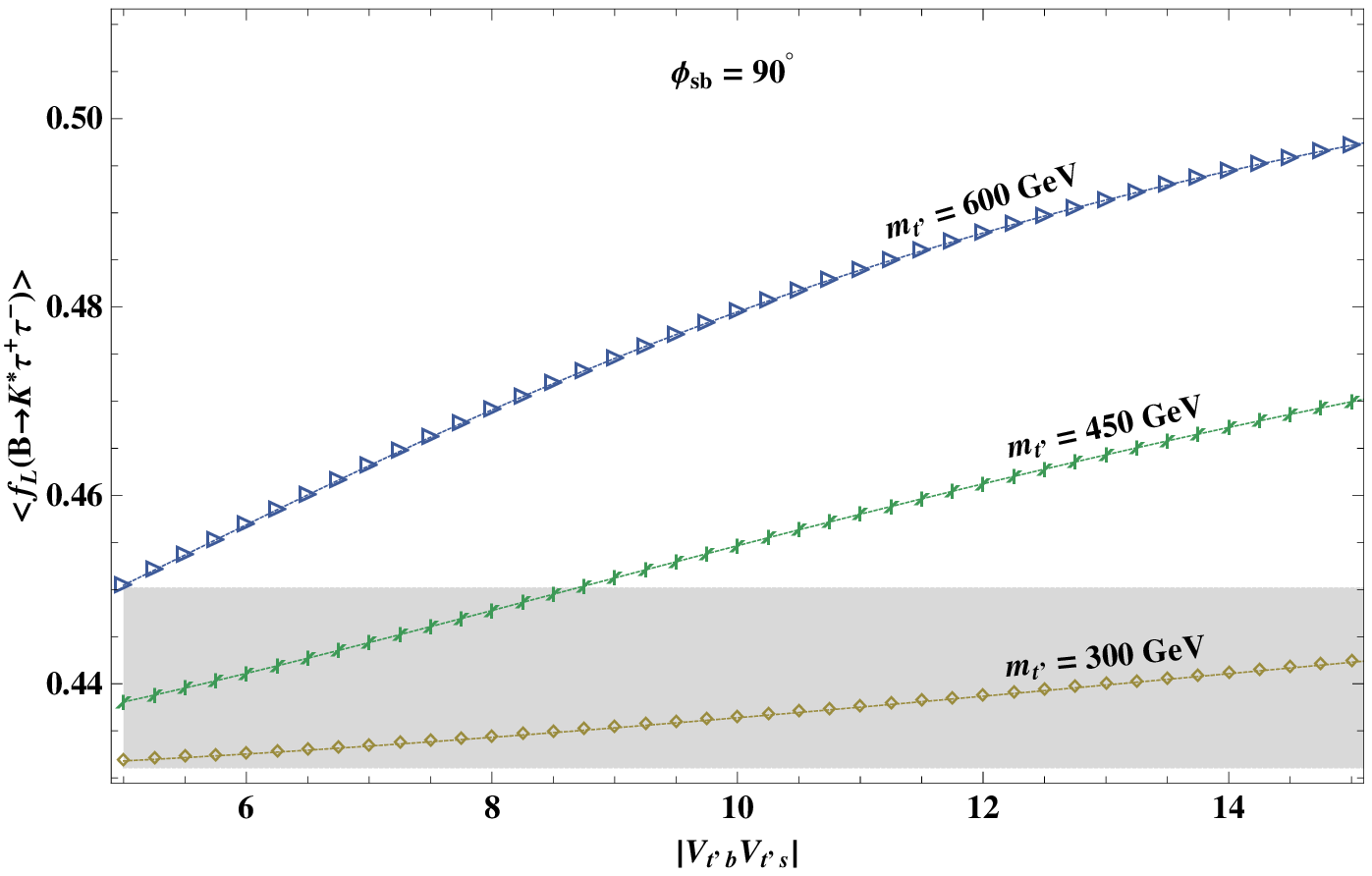} \ \ \
& \ \ \ \includegraphics[scale=0.5]{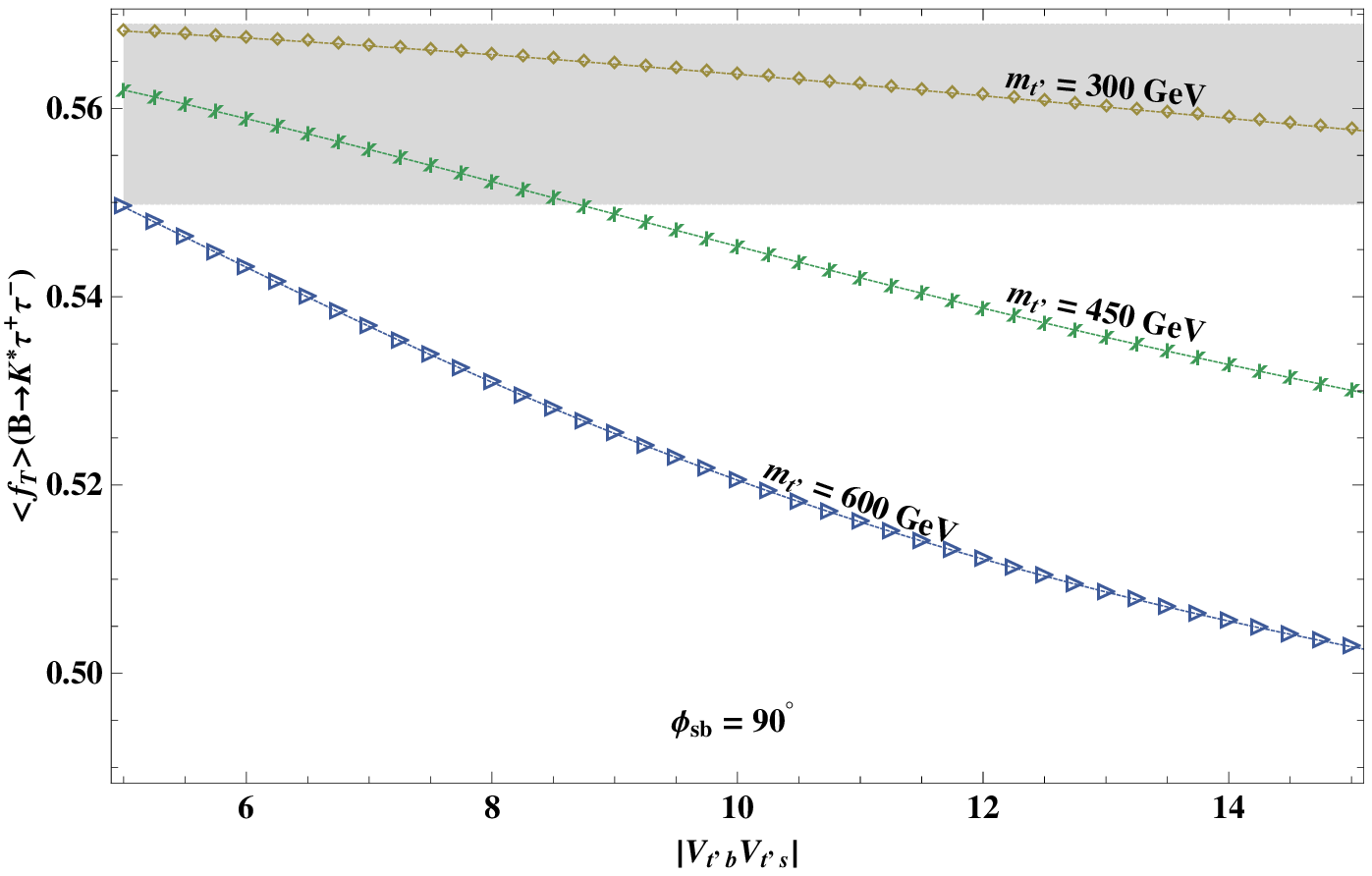}\end{tabular}
\caption{The dependence of the average longitudinal and transverse
helicity fractions for the decay $B\to K^*(892)\tau^{+}\tau^{-}$ on
$\left\vert V^{\ast}_{t^{\prime}b}V_{t^{\prime}s}\right\vert$ for
different values of $m_{t^{\prime}}$.} \label{lpt}
\end{figure*}

\section{Summary}\label{conc}

The polarization of $K^{*}$ meson in the $B\rightarrow
K^{*}\ell^{+}\ell^{-}$ ($\ell=\mu$ or $\tau$) decay is studied from
the perspective of SM4. In this respect the polarized branching
ratios $BR_{L}$, $BR_{T}$ and the helicity fractions $f_{L}, f_{T}$
of $K^{*}$ meson are studied. The explicit dependence of these
observables on the $m_{t'}, |V_{t'b}V_{t's}|$ and $\phi_{sb}$ are
also discussed. The study showed that the values of these
observables are significantly affected by changing the value of SM4
parameters. As we discussed in the numerical analysis, the polarized
branching ratios are directly proportional to the SM4 parameters
$m_{t'}$ and $|V_{t'b}V_{t's}|$ and inversely proportional to the
CKM4 phase $\phi_{sb}$ . It is found that at the maximum parametric
space of SM4 the values of $BR_{L}$ and $BR_{T}$ are enhanced up to
6-7 times of their SM values. Similarly the influence of SM4
parameters on helicity fractions $f_{L}$ and $f_{T}$ and their
average values are studied. It is shown that for the case of muons
these observables do not show any significant change in their SM
values. However, for the case of tauns the effects are quite
prominent and well distinct from their SM values. It is also noticed
that the effects of SM4 on helicity fractions are decreased when the
value of $q^{2}$ is increased and almost vanishes at the maximum
value of $q^{2}$. It is also seen that the categorical influence of
SM4 parameters $m_{t'}, |V_{t'b}V_{t's}|$ and $\phi_{sb}$ on
$\langle f_{L}\rangle$ are constructive and on $\langle
f_{T}\rangle$ are destructive. Therefore, the precise measurement of
the observables related to the polarization of $K^{*}$ meson, as
discussed in this study, not only give us an opportunity to test the
SM as well as useful to find out or put some constraint on the SM4
parameters such as $m_{t'}, |V_{t'b}V_{t's}|$ and $\phi_{sb}$. To
sum up, the precise study of the polarization of $K^{*}$ meson at
LHCb and Tevatron provide us a handy tool to dig out the status of
the extra generation of quarks.

\section*{Acknowledgments}
The authors would like to thank Professor Riazuddin, Professor
Fayyazuddin and Dr. Diego Tonelli for their valuable guidance and
helpful discussions.


\begin{thebibliography}{99}

\bibitem{Pree} Erin De Pree, G. Marshall and Marc Sher, The Fourth
Generation t-prime in Extensions of the Standard Model,arXiv:
0906.4500 [hep-ph].

\bibitem{PDG} C. Amsler \textit{et al.} [Particle Data Group], Phys. Lett.
\textbf{B667} (2008) 1.

\bibitem{Maltoni} M. Maltoni, V. A. Novikov, L. B. Okun, A. N. Rozanov and
M. I. Vysotsky, Phys. Lett. \textbf{B476} (2000) 107
[arXiv:hep-ph/9911535]; H. J. He, N. Polon- sky and S. f. Su, Phys.
Rev. \textbf{D64} (2001) 053004 [arXiv:hep-ph/0102144]; B. Holdom,
Phys. Rev. \textbf{D54} (1996) 721 [arXiv:hep-ph/9602248].

\bibitem{Kible} G. D. Kribs, T. Plehn, M. Spannowsky and T. M. P. Tait,
Phys. Rev. \textbf{D76} (2007) 075016 [arXiv:0706.3718 [hep-ph]].

\bibitem{3} W. S. Hou and C. Y. Ma, Phys. Rev. \textbf{D82} (2010) 036002.

\bibitem{4} S. Bar-Shalom, D. Oaknin and A. Soni, Phys. Rev. \textbf{D80}
(2009) 015011.

\bibitem{5} A. J. Buras, B. Duling, T. Feldmann, T. Heidsieck, C. Promberger
and S. Recksiegel, JHEP \textbf{1009} (2010) 106.

\bibitem{6} A. Soni, A. K. Alok, A. Giri, R. Mohanta and S. Nandi, Phys.
Lett. \textbf{B683} (20100 302.

\bibitem{7} O. Eberhardt, A. Lenz and J. Rohrwild, Phys. Rev. \textbf{D82}
(2010) 095006.

\bibitem{8} A. Soni, A. K. Alok, A. Giri, R. Mohanta and S. Nandi, Phys.
Rev. \textbf{D82} (2010) 033009.

\bibitem{9} A. K. Alok, A. Dighe and D. London, Phys. Rev. \textbf{D83}
(2011) 073008.

\bibitem{10} B. Holdom, Phys. Rev. Lett. \textbf{57} (1986) 2496
[Erratum-ibid. \textbf{58} (1987) 177].

\bibitem{11} C. T. Hill, M. A. Luty and E. A. Paschos, Phys. Rev. \textbf{D43%
} (1991) 3011.

\bibitem{12} T. Elliott and S. F. King, Phys. Lett. \textbf{B283} (1992) 371.

\bibitem{13} P. Q. Hung and C. Xiong, Nucl. Phys. \textbf{B848} (2011) 288.

\bibitem{14} B. Holdom, JHEP \textbf{0608} (2006) 076.

\bibitem{15} P. Q. Hung and M. Sher, Phys. Rev. \textbf{D77} (2008) 037302.

\bibitem{16} P. Q. Hung, C. Xiong, Phys. Lett. \textbf{B694} (2011) 430.

\bibitem{17} P. Q. Hung and C. Xiong, Nucl. Phys. \textbf{B847} (2011) 160.

\bibitem{18} O. Cakir, A. Senol and A. T. Tasci, Europhys. Lett. \textbf{88}
(2009) 11002.

\bibitem{19} B. Holdom, W. S. Hou, T. Hurth, M. L. Mangano, S. Sultansoy and
G. Unel, PMC Phys. \textbf{A3} (2009) 4.

\bibitem{25} T. Moroi,
Phys. Lett. \textbf{B493} (2000) 366-374, arXiv:hep-ph/0007328.

\bibitem{26} D. Chang, A. Masiero, and H. Murayama,
Phys. Rev. \textbf{D67} (2003) 075013, arXiv:hep-ph/0205111.

\bibitem{27} R. Harnik, D. T. Larson, H. Murayama, and A. Pierce,
Phys. Rev. \textbf{D69} (2004) 094024, arXiv:hep-ph/0212180.

\bibitem{28} M. Ciuchini, E. Franco, A. Masiero, and L. Silvestrini,
Phys. Rev. \textbf{D67} (2003) 075016, arXiv:hep-ph/0212397.

\bibitem{29} J. Foster, K.-i. Okumura, and L. Roszkowski, JHEP
\textbf{08} (2005) 094, arXiv:hep-ph/0506146.

\bibitem{30} M. Blanke, A. J. Buras, B. Duling, S. Gori, and A. Weiler,
JHEP \textbf{03} (2009) 001, arXiv:0809.1073 [hep-ph].

\bibitem{31} M. Blanke, A. J. Buras, B. Duling, K. Gemmler, and S. Gori, JHEP \textbf{03} (2009) 108,
arXiv:0812.3803 [hep-ph].

\bibitem{32} M. Blanke et al., JHEP \textbf{12} (2006) 003,
arXiv:hep-ph/0605214.

\bibitem{33} V. Barger et al., arXiv:0906.3745 [hep-ph].

\bibitem{bst} C.S. Kim \emph{et al.}, Phys. Lett.
B \textbf{218} (1989) 343; X.~G.~He \emph{et al.}, Phys. Rev. D
\textbf{38} (1988) 814; B. Grinstein \emph{et al.}, Nucl. Phys. B
\textbf{319} (1989) 271; N. G. Deshpande \emph{et al.}, Phys.\ Rev.\
D \textbf{39} (1989) 1461; P.~J.~O'Donnell and H.~K.~K.~Tung, Phys.\
Rev.\ D \textbf{43} (1991) 2067; N. Paver and Riazuddin, Phys. Rev.
D \textbf{45} (1992) 978; J. L. Hewett, Phys. Rev. D 53, 4964
(1996);
 T. M. Aliev, V. Bashiry, and M. Savci, Eur. Phys. J. C 35, 197
(2004); T. M. Aliev, V. Bashiry, and M. Savci, Phys. Rev. D 72,
034031 (2005); T. M. Aliev, V. Bashiry, and M. Savci, J. High Energy
Phys. 05 (2004) 037; T. M. Aliev, V. Bashiry, and M. Savci, Phys.
Rev. D 73, 034013 (2006); T. M. Aliev, V. Bashiry, and M. Savci,
Eur. Phys. J. C 40, 505 (2005); F. Kruger and L. M. Sehgal Phys.
Lett. B 380, 199 (1996); Y. G. Kim, P. Ko, and J. S. Lee, Nucl.
Phys. B544, 64 (1999); Chuan-Hung Chen and C. Q. Geng, Phys. Lett. B
516, 327 (2001); V. Bashiry, Chin. Phys. Lett. 22, 2201 (2005); W.S.
Hou, A. Soni and H. Steger, Phys. Lett. B 192, 441 (1987); W.S. Hou,
R.S. Willey and A. Soni, Phys. Rev. Lett. 58, 1608 (1987)
[Erratum-ibid. 60, 2337 (1987)]; T. Hattori, T. Hasuike and S.
Wakaizumi, Phys. Rev. D 60, 113008 (1999); T.M. Aliev, D.A. Demir
and N.K. Pak, Phys. Lett. B 389, 83 (1996); Y. Dincer, Phys. Lett.
\textbf{B 505}, 89 (2001) and references therin; C.S. Huang, W.J.
Huo and Y.L. Wu, Mod. Phys. Lett. A 14, 2453 (1999); C.S. Huang,
W.J. Huo and Y.L. Wu, Phys. Rev. D 64, 016009 (2001); A. K. Alok, et
al., JHEP \textbf{02} (2010) 053.

\bibitem{63} A. Ali, T. Mannel and T. Morosumi, Phys. Lett. B \textbf{273}, 505 (1991).
\bibitem{zp} C. W. Chiang, R. H. Li and C. D. Lu, arXiv:0911.2399.
\bibitem{colangelo} P. Colangelo, F. De Fazio, R. Feerandes and T. N. Pham, Phys.Rev. D  \textbf{74} (2006) 115006.
\bibitem{aqeel} A. Ahmed, I. Ahmed, M. A. Paracha and A. Rehman, Rev. D 84, 033010 (2011).
\bibitem{pbr} T.M. Aliev, A. Ozpineci, M. Savcı, Phys. Lett. B \textbf{511} (2001) 49.


\bibitem{LHCb} The LHCb Collaboration, CERN-LHCb-CONF-2011-038, http://cdsweb.cern.ch/record/1367849?ln=en.

\bibitem{CDF} T. Aaltonen et al., [CDF Collaboration], arXiv:1108.0695.

\bibitem{Belle} J. -T. Wei et al., [Belle Collaboration], Phys. Rev. Lett. \textbf{103}, (2009) 171801.

\bibitem{babar} B. Aubert et al., [BABAR Collaboration],  Phys. Rev. D  {\bf 79} (2009) 031102 (R).

\bibitem{Ball} A. Ali, P. Ball, L. T. Handoko and G. Hiller, Phys.
Rev. D \textbf{61}, 074024 (2000); [arXiv:hep-ph/9910221].

\bibitem{Buchalla} G. Buchalla, A. J. Buras and M. E. Lautenbacher, Rev.
Mod. Phys. \textbf{68} (1996) 1125.

\bibitem{Buras} A. J. Buras and M. Munz, Phys. Rev. \textbf{D52} (1995) 186;
A. J. Buras, M. Misiak, M. Munz and S. Pokorski, Nucl. Phys.
\textbf{B424} (1994) 374.

\bibitem{Ali} A. Ali, T. Mannel and T. Morosumi, Phys. Lett. B \textbf{273} (1991) 505.

\bibitem{Kim} C.S. Kim, T. Morozumi, A.I. Sanda, Phys. Lett. B \textbf{218}
(1989) 343.

\bibitem{Kruger} F. Kruger and L. M. Sehgal, Phys. Lett. \textbf{380} (1996)
199.

\bibitem{Grinstein} B. Grinstein, M. J. Savag and M. B. Wise, Nucl. Phys.
\textbf{B319} (1989) 271.

\bibitem{Cella} G. Cella, G. Ricciardi adn A. Vicere, Phys. Lett. \textbf{%
B258} (1991) 212.

\bibitem{Bobeth} C. Bobeth, M. Misiak and J. Urban, Nucl. Phys. \textbf{B574}
(2000) 291.

\bibitem{Asatrian} H. H. Asatrian, H. M. Asatrian, C. Grueb and M. Walker,
Phys. Lett. B507 (2001) 162.

\bibitem{Misiak} M. Misiak, Nucl. Phys. \textbf{B393} (1993) 23, Erratum,
ibid. \textbf{B439} (1995) 461.

\bibitem{Huber} T. Huber, T. Hurth, E. Lunghi, arXiv:0807.1940.



\bibitem{23} A. Saddique, M. J. Aslam and C. D. Lu, Eur. Phys. J. \textbf{C 56} (2008) 267 [arXiv:0803.0192].

\bibitem{aa1} I. Ahmed, M. A. Paracha, M. Junaid, A. Ahmed, A. Rehman and M. J. Aslam, arXiv:1107.5694.

\bibitem{pdg} K. Nakamura et al. (Particle Data Group), J. Phys. \textbf{G 37} (2010) 075021.

\bibitem{beneke} M. Beneke, T. Feldmann and D. Seidel, Nucl. Phys. \textbf{B 612} (2001) 25, [hep-ph/0106067].

\bibitem{ASoni} A.Soni et al., Phys.Rev. D \textbf{82} (2010) 033009 [arXiv:1002.0595]


\end{thebibliography}
\end{document}